\documentclass[article]{jss}

\usepackage[utf8]{inputenc} %utf8
\usepackage[T1]{fontenc}
\usepackage[english]{babel}
\usepackage{amsmath,amssymb}
\usepackage{graphicx}
\usepackage{array}
\usepackage{booktabs}

\newcolumntype{L}[1]{>{\raggedright\arraybackslash}p{#1}}

\DeclareMathSymbol{\shortminus}{\mathbin}{AMSa}{"39}
\DeclareMathSymbol{\shortplus}{\mathbin}{AMSa}{"39}
\DeclareMathOperator*{\argmax}{arg\,max}

\author{Hassan Pazira\\ARQ National Psychotrauma Center\\ The Netherlands}
\title{\pkg{BFI}: An \proglang{R} Package for Bayesian Federated Inference}
\Plainauthor{Hassan Pazira} %% comma-separated
\Plaintitle{BFI: An R Package for Bayesian Federated Inference} %% without formatting
\Shorttitle{BFI: An \proglang{R} Package} %% a short title (if necessary)

\Abstract{

Bayesian Federated Inference (BFI) estimates statistical models from
multicenter data when individual-level observations cannot be combined
across centers. We present \pkg{BFI}, an \proglang{R} package that
implements this methodology for Gaussian, binomial logistic, and survival
regression models. Each center performs a Bayesian maximum a posteriori
analysis and sends only parameter estimates and curvature information to a
central server, where these summaries are combined to approximate the
analysis of the combined data. The package supports prior specification,
structured forms of between-center heterogeneity, several parametric and
flexible baseline-hazard models for survival analysis, and treatment-effect
estimation for observational and randomized studies. We describe the
software design and the information exchanged between centers, give
reproducible workflows for these analyses, and compare the federated results
with pooled-data analyses for Gaussian, logistic, and survival models.

}
\Plainkeywords{Bayesian inference, decentralized data, federated inference, distributed inference, survival analysis, R, BFI} %% without formatting
\Address{
  Hassan Pazira\\
ARQ Center'45\\
ARQ National Psychotrauma Center, Diemen, The Netherlands\\
E-mail: \email{h.pazira@arq.org}\\
URL: \url{https://radboudumc-biostatistics.github.io/people/hassan/} \\
}

\begin{document}

%% include your article here, just as usual
%% Note that you should use the \pkg{}, \proglang{} and \code{} commands.

\section{Introduction}
\label{sec:intro}

Statistical analyses based on data collected in multiple centers can benefit
substantially from the larger effective sample size obtained by combining
information across centers. In many applications, however, individual-level
data cannot readily be pooled because of regulatory, governance, privacy, or
logistical constraints. This is particularly problematic when the sample
sizes in the individual centers are small relative to the number of model
parameters or, for time-to-event outcomes, when the number of observed events
is limited. Such settings may lead to unstable estimation and overfitting and
motivate statistical methods that combine information across centers without
requiring the underlying individual-level data to be transferred
\citep{Jonker2024BFI,Jonker2025BFI,Pazira2026BFISurvival}.

Federated learning (FL) provides a general framework for learning from
decentralized data while keeping the original data at their local sources.
A widely used implementation of FL repeatedly exchanges locally computed
model updates between participating centers and a central server until a
convergence criterion is reached \citep{McMahan2017}. Related approaches have
also been developed in the field of distributed statistical inference, where
local estimates or likelihood-based summaries are combined to approximate
inference based on the complete data set \citep{Jordan2019}. For conventional
statistical analyses involving multiple collaborating centers, repeated
communication and repeated local model fitting may, however, be inconvenient.
Moreover, accurate quantification of statistical uncertainty and the ability
to account for heterogeneity between centers are important requirements in
many applications.

Several software implementations for analyzing decentralized health data are
already available. \pkg{DataSHIELD} \citep{DataSHIELD} provides an
infrastructure in which analysis commands are executed on remote servers and
only non-disclosive summaries are returned, and the \proglang{R} package
\pkg{distcomp} \citep{distcomp} implements distributed fitting of a stratified
Cox model and a distributed singular value decomposition across participating
sites. Closest in spirit to the present work is the \proglang{R} package
\pkg{pda} \citep{pdapackage}, which implements one-shot and few-shot
distributed algorithms for linear, logistic, Poisson and Cox regression based
on surrogate likelihood functions \citep{Duan2020,Luo2022}. Section~\ref{sec:discussion}
discusses how \pkg{BFI} relates to these implementations.

Bayesian Federated Inference (BFI) was developed as a one-shot approach for
this setting \citep{Jonker2024BFI}. Suppose that data are stored in $L$
separate centers and cannot be merged. Each center performs a Bayesian
analysis of its local data and computes the maximum a posteriori (MAP)
estimate of the model parameters together with the curvature of the local
log-posterior density at the MAP estimate. These inferential summaries, rather
than the individual-level observations, are transferred to a central server.
Using a second-order approximation of the local log-posterior densities, the
local results can then be combined to approximate the MAP estimate and its
curvature that would have been obtained from the fictive combined data set.
Consequently, the local analyses need to be carried out only once, and no
iterative exchange of updated model parameters between the central server and
the local centers is required \citep{Jonker2024BFI,Jonker2025BFI}.

The statistical methodology underlying BFI has been developed in a series of
papers. The original framework was introduced for parametric statistical
models, with particular attention to generalized linear models (GLMs)
\citep{Jonker2024BFI}. It was subsequently extended to accommodate several
forms of heterogeneity across centers, including center-specific intercepts,
regression parameters and nuisance parameters, clustering of centers, and
center-specific covariates \citep{Jonker2025BFI}. The methodology was also
extended to time-to-event outcomes, including estimation of regression
parameters and parameters describing the baseline hazard function
\citep{Pazira2026BFISurvival}. These publications provide the mathematical
derivations, asymptotic results, simulation studies, and substantive
applications of the BFI methodology. The purpose of the present article is
therefore not to repeat these methodological developments, but to present
their software implementation and provide a unified practical framework for
applying BFI.

The methodology is implemented in the \proglang{R} \citep{R} package \pkg{BFI}
\citep{BFIpackage}, available from the Comprehensive \proglang{R} Archive Network
(CRAN). Version~3.2.0 of the package implements BFI for Gaussian, binomial
(logistic), and survival regression models. For survival analysis, the
implemented baseline hazard specifications include exponential, Weibull,
Gompertz, exponentiated polynomial, piecewise exponential, and an unspecified
baseline hazard corresponding to Cox proportional hazards regression. The
package also provides functionality for stratified BFI analyses to accommodate
several forms of between-center heterogeneity. In addition, functionality is available for estimating treatment effects in Gaussian, binomial, and survival models. For survival outcomes, treatment-effect estimation is implemented using a Cox proportional hazards model with an unspecified baseline hazard and a weighted partial log-likelihood
\citep{BFIpackage}.

The design of \pkg{BFI} reflects the two distinct stages of a federated
analysis. At each local center, the function \code{MAP.estimation()} fits the
specified model and returns the quantities required for federated
aggregation, most importantly the local MAP estimate and the corresponding
curvature matrix. Prior precision matrices can be constructed using
\code{inv.prior.cov()}. At the central server, the function \code{bfi()}
combines the local inferential results and returns the BFI estimates,
posterior standard deviations, and the aggregated curvature matrix. This
separation between local estimation and central aggregation makes the data
flow explicit and allows the complete federated analysis to be reproduced
without constructing a pooled individual-level data set.

This article focuses on the statistical software and its use in practice.
First, we give a concise summary of the BFI methodology required to understand
the implementation. We then describe the design of the \pkg{BFI} package, its
main functions, input and output objects, prior specification, and the
information exchanged between local centers and the central server. Worked
examples are then presented for Gaussian regression, between-center
heterogeneity, survival models, and treatment-effect estimation. Numerical
validation examples subsequently compare BFI with corresponding pooled-data
analyses for Gaussian, logistic, and survival models when pooling is available
for validation purposes. Finally, the scope and current limitations of the
software are discussed.

\section{Bayesian Federated Inference}
\label{sec:bfi_method}

This section briefly summarizes the Bayesian Federated Inference (BFI)
framework required to understand the implementation in \pkg{BFI}. Detailed derivations, theoretical properties, simulation studies, and extensions are
available in the methodological papers
\citep{Jonker2024BFI,Jonker2025BFI,Pazira2026BFISurvival}.

\subsection{Setting and local inference}
\label{sec:bfi_local}

Suppose that data are available in $L$ centers but cannot be combined into a
single data set. Let

\begin{equation}
  \mathcal{D}_{\ell}
  =
  \left\{
  (\mathbf{x}_{\ell 1}, y_{\ell 1}),
  \ldots,
  (\mathbf{x}_{\ell n_{\ell}}, y_{\ell n_{\ell}})
  \right\},
  \qquad \ell = 1,\ldots,L,
\end{equation}

denote the data in center $\ell$, where $n_{\ell}$ is the local sample size.
The fictive combined data set is

\begin{equation}
  \mathcal{D}
  =
  \bigcup_{\ell=1}^{L}\mathcal{D}_{\ell},
  \qquad
  n = \sum_{\ell=1}^{L} n_{\ell}.
\end{equation}

The data set $\mathcal{D}$ is introduced only to define the inferential target;
it does not need to be physically constructed. The objective of BFI is to
approximate the inference that would have been obtained from
$\mathcal{D}$ by using only inference results calculated separately in the
$L$ centers \citep{Jonker2024BFI}.

Let $\boldsymbol{\theta}\in\mathbb{R}^{d}$ denote the parameter vector of the
statistical model. The parameters can contain regression coefficients as well
as nuisance parameters or parameters defining a baseline hazard function,
depending on the model under consideration
\citep{Jonker2024BFI,Pazira2026BFISurvival}. For the fictive combined data set,
let $p(\boldsymbol{\theta})$ denote the prior density, while
$p_{\ell}(\boldsymbol{\theta})$ denotes the prior density used in center
$\ell$. The corresponding posterior densities satisfy

\begin{equation}
  p(\boldsymbol{\theta}\mid\mathcal{D})
  \propto
  \frac{p(\boldsymbol{\theta})}
       {\prod_{\ell=1}^{L}p_{\ell}(\boldsymbol{\theta})}
  \prod_{\ell=1}^{L}
  p_{\ell}(\boldsymbol{\theta}\mid\mathcal{D}_{\ell}).
  \label{eq:posterior_decomposition}
\end{equation}

Thus, the posterior density corresponding to the fictive combined data can be
expressed in terms of the local posterior densities and the prior
distributions \citep{Jonker2024BFI}.

In the implementation considered here, zero-mean Gaussian priors are used.
Let

\begin{equation}
  p_{\ell}(\boldsymbol{\theta})
  =
  \mathcal{N}
  \left(
    \mathbf{0},
    \boldsymbol{\Lambda}_{\ell}^{-1}
  \right),
  \qquad
  p(\boldsymbol{\theta})
  =
  \mathcal{N}
  \left(
    \mathbf{0},
    \boldsymbol{\Lambda}^{-1}
  \right),
  \label{eq:bfi_priors}
\end{equation}

where $\boldsymbol{\Lambda}_{\ell}$ and
$\boldsymbol{\Lambda}$ are the prior precision matrices for center $\ell$
and the fictive combined analysis, respectively. 
Parameters that are constrained to be positive are handled on a transformed
scale. 
For parameters constrained to be positive, the Gaussian prior and the
quadratic approximation can be formulated on a transformed scale
\citep{Jonker2025BFI}. In the Gaussian implementation of \pkg{BFI}, the
dispersion parameter is internally parameterized as
\(\eta=\log(\sigma^2)\). Estimation and the corresponding curvature
calculations are performed on this scale, whereas the reported
\code{theta_hat} value is back-transformed to the residual variance
\(\sigma^2=\exp(\eta)\) \citep{BFIpackage}.
The local and combined prior distributions do not have to be identical \citep{Jonker2024BFI}.

In each center, the local maximum a posteriori (MAP) estimator is

\begin{equation}
  \widehat{\boldsymbol{\theta}}_{\ell}
  =
  \argmax_{\boldsymbol{\theta}}
  p_{\ell}
  \left(
    \boldsymbol{\theta}\mid\mathcal{D}_{\ell}
  \right).
  \label{eq:local_map}
\end{equation}

In addition to the MAP estimate, BFI uses information about the local
curvature of the log-posterior density. Define

\begin{equation}
  \widehat{\mathbf{A}}_{\ell}
  =
  -
  \left.
  \nabla_{\boldsymbol{\theta}}^{2}
  \log
  p_{\ell}
  \left(
    \boldsymbol{\theta}\mid\mathcal{D}_{\ell}
  \right)
  \right|_{
    \boldsymbol{\theta}
    =
    \widehat{\boldsymbol{\theta}}_{\ell}
  }.
  \label{eq:local_curvature}
\end{equation}

The matrix $\widehat{\mathbf{A}}_{\ell}$ is therefore minus the Hessian of the
local log-posterior density evaluated at the local MAP estimate. Around
$\widehat{\boldsymbol{\theta}}_{\ell}$, the local log-posterior density is
approximated to second order by

\begin{equation}
  \log
  p_{\ell}
  \left(
    \boldsymbol{\theta}\mid\mathcal{D}_{\ell}
  \right)
  \approx
  \log
  p_{\ell}
  \left(
    \widehat{\boldsymbol{\theta}}_{\ell}
    \mid
    \mathcal{D}_{\ell}
  \right)
  -
  \frac{1}{2}
  \left(
    \boldsymbol{\theta}
    -
    \widehat{\boldsymbol{\theta}}_{\ell}
  \right)^{\top}
  \widehat{\mathbf{A}}_{\ell}
  \left(
    \boldsymbol{\theta}
    -
    \widehat{\boldsymbol{\theta}}_{\ell}
  \right).
  \label{eq:local_quadratic}
\end{equation}

This quadratic approximation is the central ingredient of the BFI
aggregation rule \citep{Jonker2024BFI}.

\subsection{Central BFI aggregation}
\label{sec:bfi_aggregation}

After the local analyses have been completed, center $\ell$ provides the
quantities
$\widehat{\boldsymbol{\theta}}_{\ell}$,
$\widehat{\mathbf{A}}_{\ell}$, and the information required to specify
$\boldsymbol{\Lambda}_{\ell}$. Individual-level observations are not needed
for the subsequent BFI aggregation \citep{Jonker2024BFI}.

Combining the quadratic approximations in
Equation~\ref{eq:local_quadratic} with the Gaussian prior distributions in
Equation~\ref{eq:bfi_priors} gives the BFI approximation to the curvature
matrix of the posterior distribution corresponding to the fictive combined
data set:

\begin{equation}
  \widehat{\mathbf{A}}_{\mathrm{BFI}}
  =
  \sum_{\ell=1}^{L}
  \widehat{\mathbf{A}}_{\ell}
  +
  \boldsymbol{\Lambda}
  -
  \sum_{\ell=1}^{L}
  \boldsymbol{\Lambda}_{\ell}.
  \label{eq:A_bfi}
\end{equation}

The corresponding BFI estimator of the model parameter vector is

\begin{equation}
  \widehat{\boldsymbol{\theta}}_{\mathrm{BFI}}
  =
  \widehat{\mathbf{A}}_{\mathrm{BFI}}^{-1}
  \sum_{\ell=1}^{L}
  \widehat{\mathbf{A}}_{\ell}
  \widehat{\boldsymbol{\theta}}_{\ell}.
  \label{eq:theta_bfi}
\end{equation}

Equations~\ref{eq:A_bfi} and~\ref{eq:theta_bfi} are the core aggregation
formulae implemented by \pkg{BFI}. They approximate the MAP estimate and the
curvature that would have been obtained from an analysis of the fictive
combined data set \citep{Jonker2024BFI}.

On the parameterization used to define the curvature matrix, an approximate
covariance matrix for the BFI estimator is obtained from

\begin{equation}
  \widehat{\mathrm{Var}}
  \left(
    \widehat{\boldsymbol{\theta}}_{\mathrm{BFI}}
  \right)
  =
  \widehat{\mathbf{A}}_{\mathrm{BFI}}^{-1}.
  \label{eq:cov_bfi}
\end{equation}

Consequently, for a parameter reported on the same scale as that used to
define the curvature matrix, the approximate posterior standard deviation of
the \(k\)th parameter is

\begin{equation}
  \widehat{\mathrm{sd}}
  \left(
    \widehat{\theta}_{\mathrm{BFI},k}
  \right)
  =
  \sqrt{
    \left[
      \widehat{\mathbf{A}}_{\mathrm{BFI}}^{-1}
    \right]_{kk}
  }.
  \label{eq:sd_bfi}
\end{equation}

If the reported parameter is a transformation of the parameterization used
to define the curvature matrix, the corresponding
uncertainty is transformed using the delta method. In particular, for the
Gaussian residual variance, the working parameter is
\(\eta=\log(\sigma^2)\), whereas the reported parameter is
\(\sigma^2=\exp(\eta)\). Thus,

\[
  \widehat{\mathrm{sd}}
  \left(
    \widehat{\sigma}^{2}
  \right)
  \approx
  \widehat{\sigma}^{2}
  \sqrt{
    \left[
      \widehat{\mathbf{A}}_{\mathrm{BFI}}^{-1}
    \right]_{\eta\eta}
  }.
\]

Approximate credible intervals are constructed using the Gaussian
approximation on the scale on which the curvature is defined. For the
Gaussian residual variance, the interval is therefore first constructed on
the \(\log(\sigma^2)\) scale and then back-transformed to the
\(\sigma^2\) scale:

\[
  \left[
    \widehat{\sigma}^{2}
    \exp\left\{-z_{0.975}\,s_{\eta}\right\},
    \;
    \widehat{\sigma}^{2}
    \exp\left\{z_{0.975}\,s_{\eta}\right\}
  \right],
  \qquad
  s_{\eta}
  =
  \sqrt{
    \left[
      \widehat{\mathbf{A}}_{\mathrm{BFI}}^{-1}
    \right]_{\eta\eta}
  }.
\]
where \(z_{0.975}\) denotes the 97.5th percentile of the standard normal
distribution.
Consequently, the reported credible interval for \(\sigma^2\) is generally
asymmetric on the original scale.

A practical feature of Equations~\ref{eq:A_bfi} and~\ref{eq:theta_bfi} is that
the central aggregation requires a single set of inference results from each
center. Once the local MAP estimates and curvature matrices have been
obtained, repeated refitting of the local models is not required for the
standard BFI aggregation \citep{Jonker2024BFI,Jonker2025BFI}.

\subsection{Heterogeneity across centers}
\label{sec:bfi_heterogeneity}

Equations~\ref{eq:A_bfi} and~\ref{eq:theta_bfi} describe the basic setting in
which the model parameters represented by
$\boldsymbol{\theta}$ are shared across centers. In applications, however,
some model parameters may differ between centers. The BFI framework has been
extended to distinguish parameters that are common across centers from
parameters that are center-specific \citep{Jonker2025BFI}.

For regression models, the extensions include heterogeneity in outcome means
through center-specific intercepts, center-specific regression effects,
center-specific nuisance parameters, clustering of centers, and
center-specific covariates \citep{Jonker2025BFI}. In survival models,
heterogeneity can similarly be represented by allowing selected parameters of
the baseline hazard function to vary across centers
\citep{Pazira2026BFISurvival}. In these settings, the dimensions and structure
of the aggregated parameter vector and curvature matrix are modified so that
shared parameters are combined across centers while the relevant
center-specific parameters remain distinct
\citep{Jonker2025BFI,Pazira2026BFISurvival}.

Under the regularity conditions described by \citet{Jonker2025BFI}, the BFI estimators are asymptotically efficient in both the homogeneous setting and the considered heterogeneous settings. Detailed derivations of the heterogeneous aggregation formulae are therefore omitted here. The remainder of this article focuses on how the corresponding
methodology is exposed to the user through the functions and arguments of the
\pkg{BFI} package.

\bigskip

\section{Software design and implementation}
\label{sec:software}

The \pkg{BFI} package implements the statistical computations required on
both sides of a Bayesian Federated Inference analysis: local model fitting in
the participating centers and aggregation of the local inference results at a
central server. The current CRAN version of the package is version~3.2.0.
It requires \proglang{R} version~3.5.0 or later, imports functionality from
the \pkg{stats} package, and does not require compiled code
\citep{BFIpackage}.

The implementation follows the statistical structure described in
Section~\ref{sec:bfi_method}. In a standard BFI analysis, each center fits the
agreed statistical model using only its local individual-level data. The
resulting MAP estimate and the corresponding curvature matrix are then
provided to the central analysis. The central server combines these local
inferential summaries using the BFI aggregation equations rather than
refitting the model to individual-level observations
\citep{Jonker2024BFI,Jonker2025BFI}.

\subsection{Computational architecture}
\label{sec:architecture}

The package separates a BFI analysis into a local stage and a central stage.
For a standard analysis, the workflow consists of the following steps:

\begin{enumerate}
  \item The statistical model, covariates, parameterization, and prior
  specification are agreed upon across the participating centers.

  \item In each center $\ell$, a prior precision matrix
  $\boldsymbol{\Lambda}_{\ell}$ is specified and the local model is fitted
  using \code{MAP.estimation()}. The main inferential outputs are the local
  MAP estimate $\widehat{\boldsymbol{\theta}}_{\ell}$ and the corresponding
  minus-curvature matrix $\widehat{\mathbf{A}}_{\ell}$.

  \item At the central server, the local estimates, curvature matrices, and
  prior information are supplied to \code{bfi()}, which computes the
  aggregated BFI estimate and its curvature matrix.
\end{enumerate}

Thus, for standard model estimation, individual-level response and covariate
data are used by \code{MAP.estimation()} within the local centers but are not
arguments of \code{bfi()} at the central server. The central calculation is
instead based on the inferential summaries produced by the local analyses
\citep{Jonker2024BFI,BFIpackage}.

The package does not require the prior used for the fictive combined analysis
to be identical to the priors used in the local centers. Consequently, the
central BFI calculation keeps track of both the local prior precision
matrices $\boldsymbol{\Lambda}_{\ell}$ and the prior precision matrix
$\boldsymbol{\Lambda}$ associated with the fictive combined data set. When
the same prior is used in all centers, the interface allows the corresponding
input to be specified more compactly \citep{BFIpackage}.

The local models must represent compatible parameterizations of the intended
statistical model. In particular, corresponding covariates must have the same
meaning, coding, reference categories, and resulting parameter names across
centers. For robustness and reproducibility, using a common covariate ordering
across centers is recommended. 
The central \code{bfi()} function can align local \code{theta_hat} vectors,
\code{A_hat} matrices, and the corresponding prior precision matrices by
parameter name when the same set of parameters is supplied in different
orders. Nevertheless, consistent model specification remains important to
ensure that corresponding parameters have the same statistical meaning across
centers \citep{BFIpackage}.

\subsection[Prior specification with inv.prior.cov()]
{Prior specification with \code{inv.prior.cov()}}
\label{sec:prior_implementation}

The BFI methodology implemented in the package uses zero-mean Gaussian priors,
parameterized through their inverse covariance, or precision, matrices. The
utility function \code{inv.prior.cov()} facilitates construction of these
matrices. Its basic syntax is

\begin{Code}
inv.prior.cov(X, lambda = 1, family = "gaussian")
\end{Code}

where \code{X} contains the covariates and \code{lambda} determines the
diagonal elements of the prior precision matrix. The required dimension of
the matrix depends on the model family, the number and type of covariates,
whether an intercept is fitted, and the additional parameters required by the
model.

Categorical covariates require more than one regression parameter when they
have more than two levels. The function accounts for this expansion when
determining the dimension of the prior matrix. For Gaussian regression, an
additional parameter is included for the residual variance. For survival
models, the required dimension additionally depends on the number of
parameters used to describe the baseline hazard function. The same function
can construct the larger prior precision matrices required for stratified BFI
models in which selected parameters are allowed to vary across centers
\citep{BFIpackage}.

For example, a common diagonal prior precision matrix for a Gaussian model can
be constructed using

\begin{Code}
Lambda <- inv.prior.cov(X, lambda = 0.01, family = "gaussian")
\end{Code}

where a value of \code{lambda = 0.01} corresponds to a prior variance of
$100$ for parameters to which that precision value is assigned.

\subsection[Local estimation with MAP.estimation()]
{Local estimation with \code{MAP.estimation()}}
\label{sec:map_implementation}

The function \code{MAP.estimation()} performs the local Bayesian analysis. Its
principal arguments are the response \code{y}, the covariate data \code{X},
the model \code{family}, and the prior precision matrix \code{Lambda}. A
typical Gaussian analysis within one center takes the form

\begin{Code}
fit_local <- MAP.estimation(y = y, X = X, family = "gaussian", 
                            Lambda = Lambda)
\end{Code}

For binomial models, the package uses the logit link. For Gaussian models, the
identity link is used and the residual variance is included among the model
parameters. When \code{family = "survival"}, the model specification is
completed by selecting the baseline hazard through the \code{basehaz}
argument \citep{BFIpackage}.

The returned object contains, among other quantities,

\begin{itemize}
  \item \code{theta\_hat}: the vector of local MAP estimates;
  \item \code{A\_hat}: minus the curvature matrix of the local log-posterior
  density evaluated at the MAP estimate;
  \item \code{sd}: posterior standard deviations obtained from
  $\sqrt{\mathrm{diag}(\widehat{\mathbf{A}}_{\ell}^{-1})}$;
  \item \code{Lambda}: the prior precision matrix used in the analysis;
  \item model information such as the parameter names, formula, local sample
  size, model family, and convergence information.
\end{itemize}

The quantities \code{theta\_hat} and \code{A\_hat} correspond directly to
$\widehat{\boldsymbol{\theta}}_{\ell}$ and
$\widehat{\mathbf{A}}_{\ell}$ in
Equations~\ref{eq:local_map} and~\ref{eq:local_curvature}. This direct
correspondence between the mathematical notation and the software objects is
useful when constructing a federated workflow.

\subsection[Central aggregation with bfi()]
{Central aggregation with \code{bfi()}}
\label{sec:bfi_implementation}

After all local analyses have been completed, the corresponding inferential
results are collected at the central server. For $L$ centers, the local MAP
estimates and curvature matrices are stored in lists:

\begin{Code}
theta_hats <- list(fit1$theta_hat, fit2$theta_hat, ...)
A_hats     <- list(fit1$A_hat, fit2$A_hat, ...)
\end{Code}

For a standard homogeneous analysis, the central calculation can then be
performed using

\begin{Code}
fit_bfi <- bfi(theta_hats = theta_hats, A_hats = A_hats, Lambda = Lambda,
               family = "gaussian")
\end{Code}

When different priors are used in the local centers, the argument
\code{Lambda} can instead contain the local prior precision matrices together
with the precision matrix selected for the fictive combined analysis. The
ordering of center-specific inputs must be consistent across the lists
supplied to \code{bfi()} \citep{BFIpackage}.

The principal output elements of \code{bfi()} are

\begin{itemize}
  \item \code{theta\_hat}: the aggregated BFI parameter estimates;
  \item \code{A\_hat}: minus the curvature matrix for the aggregated model;
  \item \code{sd}: posterior standard deviations computed from the inverse of
  the aggregated curvature matrix.
\end{itemize}

These quantities correspond to
$\widehat{\boldsymbol{\theta}}_{\mathrm{BFI}}$,
$\widehat{\mathbf{A}}_{\mathrm{BFI}}$, and the standard deviations defined in
Equations~\ref{eq:theta_bfi}, \ref{eq:A_bfi}, and~\ref{eq:sd_bfi},
respectively.

Both \code{MAP.estimation()} and \code{bfi()} return objects of class
\code{"bfi"}, for which the package provides the S3 methods \code{print()},
\code{summary()}, \code{coef()}, and \code{vcov()}. The \code{coef()} method
returns the estimates as a named vector. The \code{vcov()} method returns the
approximate posterior covariance matrix in the parameterization returned by
\code{coef()}; for Gaussian residual-variance parameters, the covariance
matrix is transformed from the \(\log(\sigma^2)\) scale to the original
\(\sigma^2\) scale using the multivariate delta method
\citep{BFIpackage}. For example,

\begin{Code}
summary(fit_bfi)
\end{Code}

reports the parameter estimates, posterior standard deviations, and
approximate 95\% credible intervals. The curvature matrix can additionally be
printed using

\begin{Code}
summary(fit_bfi, cur_mat = TRUE)
\end{Code}

\subsection{Supported model classes}
\label{sec:model_classes}

Version~3.2.0 of \pkg{BFI} implements Gaussian, binomial, and survival
regression models. Table~\ref{tab:model_classes} summarizes the available
model specifications.

\begin{table}[t!]
\centering
\begin{tabular}{@{}p{0.12\textwidth}p{0.2\textwidth}p{0.61\textwidth}@{}}
\toprule
\code{family} &
Model &
Implementation \\
\midrule

\code{"gaussian"} &
Linear regression &
Identity link; regression coefficients and residual variance are estimated. \\

\code{"binomial"} &
Logistic regression &
Logit link; binary or grouped-binomial responses can be analyzed. \\

\code{"survival"} &
Survival regression &
Supports exponential, Weibull, Gompertz, exponentiated polynomial,
piecewise exponential, and unspecified baseline hazard functions. \\

\bottomrule
\end{tabular}
\caption{Statistical model classes implemented in \pkg{BFI}.}
\label{tab:model_classes}
\end{table}

For survival models, the baseline hazard is selected through the
\code{basehaz} argument. The available choices are
\code{"exp"}, \code{"weibul"}, \code{"gomp"}, \code{"poly"},
\code{"pwexp"}, and \code{"unspecified"}. The last option corresponds to a
semi-parametric Cox proportional hazards analysis \citep{Cox72} in which the regression
coefficients are estimated from the partial log-likelihood
\citep{Pazira2026BFISurvival,BFIpackage}.

The parametric survival models estimate both regression coefficients and
parameters describing the baseline hazard. Consequently, the number and
interpretation of parameters passed between the local centers and central
server depend on the selected baseline-hazard specification. The package
handles these differences internally through the arguments of
\code{inv.prior.cov()}, \code{MAP.estimation()}, and \code{bfi()}
\citep{Pazira2026BFISurvival,BFIpackage}.

\subsection{Main functions and supporting utilities}
\label{sec:function_overview}

Table~\ref{tab:bfi_functions} gives an overview of the functions that are most
relevant for a BFI workflow. The package documentation identifies
\code{MAP.estimation()} and \code{bfi()} as the two main functions, while the
remaining functions provide supporting functionality \citep{BFIpackage}. The
second column indicates where each function is used, which makes the
separation between local estimation and central aggregation described in
Section~\ref{sec:architecture} visible in the interface itself.

\begin{table}[t!]
\centering
\begin{tabular}{@{}L{0.2\textwidth}L{0.19\textwidth}L{0.54\textwidth}@{}}
\toprule
Function &
Role &
Purpose \\
\midrule

\code{MAP.estimation()} &
Local &
Computes local MAP estimates, curvature matrices, posterior standard
deviations, and model-specific information. \\

\code{bfi()} &
Central &
Combines local inference results into the aggregated BFI model. \\

\code{summary()} &
Local and central &
Summarizes objects returned by \code{MAP.estimation()} or \code{bfi()},
including estimates, standard deviations, and credible intervals. \\

\code{print()} &
Local and central &
Prints the estimates of objects returned by \code{MAP.estimation()} or
\code{bfi()}. \\

\code{coef()} &
Local and central &
Extracts the estimates as a named vector. \\

\code{vcov()} &
Local and central &
Returns the approximate posterior covariance matrix on the parameter scale
returned by \code{coef()}, using a delta-method transformation for Gaussian
residual-variance parameters. \\

\code{inv.prior.cov()} &
Local and central &
Constructs diagonal Gaussian prior precision matrices with dimensions
appropriate for the selected model, including the enlarged matrices required
for stratified models. \\

\code{hazards.fun()} &
Auxiliary &
Evaluates the baseline hazard, cumulative hazard and survival functions of a
fitted parametric survival model, and the corresponding quantities for a new
covariate vector. \\

\code{surv.simulate()} &
Auxiliary &
Generates right-censored survival data with a prespecified censoring rate, for
simulation and illustration. \\

\code{n.par()} &
Auxiliary &
Reports the number and names of the regression coefficients implied by a set
of covariates, together with the number of predictors and observations. \\

\code{b.diag()} &
Auxiliary &
Constructs block-diagonal matrices, which are useful when assembling prior
precision matrices for stratified models. \\

\bottomrule
\end{tabular}
\caption{Main functions and selected supporting functions in \pkg{BFI}.}
\label{tab:bfi_functions}
\end{table}

With the exception of \code{hazards.fun()}, which is used in
Section~\ref{sec:survival}, the functions marked as auxiliary in
Table~\ref{tab:bfi_functions} are not required for a standard BFI analysis and
are not illustrated further in this article. Their arguments and return values are documented in the package manual \citep{BFIpackage}. The package also
exports a small number of lower-level functions that are used internally by
\code{MAP.estimation()} and \code{bfi()}. These are documented under
\code{BFI-internal} and are not intended for direct use; an example is \code{ql.LRT()}, which \code{MAP.estimation()} uses
to select the polynomial order when \code{basehaz = "poly"}.

\subsection{Non-standard information flows}
\label{sec:nonstandard_flow}

The standard workflow described above is based on the local
\code{theta\_hat} and \code{A\_hat} objects. Some of the more advanced
features of the package require additional summary information.

For the exponentiated polynomial baseline hazard, local model selection may
lead to different selected polynomial orders across centers. In this case,
\code{MAP.estimation()} can return \code{q\_l} and
\code{theta\_A\_poly}, containing information for multiple candidate
polynomial orders. These objects can subsequently be supplied to
\code{bfi()} through \code{q\_ls} and \code{theta\_A\_polys}, allowing the
central analysis to construct a compatible survival model without access to
the individual survival observations \citep{Pazira2026BFISurvival,BFIpackage}.

Similarly, stratified analyses require information indicating which
parameters or center characteristics are allowed to vary. The arguments
\code{stratified}, \code{strat\_par}, and \code{center\_spec} control these
extensions. Additional metadata can also be supplied when a categorical
covariate has no observations in one of its categories within a local center
\citep{Jonker2025BFI,BFIpackage}.

Finally, treatment-effect estimation is a distinct extension of the standard
one-shot model-fitting workflow. For observational data, the implementation
uses a first round to estimate a federated propensity-score model and a
second round to estimate the treatment effect using the resulting propensity
information. For randomized trials with known treatment probabilities, the
propensity-score estimation round is not required. This functionality is described separately in Section~\ref{sec:treatment_effect}.

\section{A complete BFI workflow}
\label{sec:workflow}

We now illustrate a complete BFI analysis using the \code{Nurses} data
included in the package. The purpose of this section is primarily to
demonstrate the software workflow: construction of the local analyses,
extraction of the quantities that are communicated to the central server,
central aggregation with \code{bfi()}, and comparison with an analysis of the
pooled data for validation.

The \code{Nurses} data originate from a simulated multilevel data set for a
hypothetical study of job-related stress among nurses working in different
hospitals \citep{Jonker2025BFI}. The data contain 1,000 nurses from 25
hospitals. The number of nurses per hospital ranges from 36 to 52.
Information is available on job-related stress, age, gender, years of
experience, ward type, hospital, and hospital size
\citep{Jonker2025BFI}.

In a real federated analysis, the complete \code{Nurses} object would not be
available at one location. Each hospital would have access only to its own
data. For reproducibility of the present article, however, the complete data
set is available in \pkg{BFI}. We therefore mimic the decentralized setting
by splitting the data according to hospital. None of the pooled individual
data created for this demonstration are required by the BFI aggregation
itself.

\subsection{Loading and preparing the data}
\label{sec:workflow_data}

The package and data are loaded as follows.

\begin{CodeInput}
R> library("BFI")
R> data("Nurses", package = "BFI")
\end{CodeInput}

We consider a Gaussian linear regression model for job-related stress with
age, gender, experience, and ward type as covariates,

\begin{equation}
  \mathrm{stress}_{\ell i}
  =
  \beta_0
  +
  \beta_1 \mathrm{age}_{\ell i}
  +
  \beta_2 \mathrm{gender}_{\ell i}
  +
  \beta_3 \mathrm{experience}_{\ell i}
  +
  \beta_4 \mathrm{wardtype}_{\ell i}
  +
  \varepsilon_{\ell i},
  \label{eq:nurses_model}
\end{equation}

where $\varepsilon_{\ell i} \sim \mathcal{N}(0,\sigma^2)$, $\ell$ denotes the hospital and $i$ the nurse within that hospital.
This model is also used in the methodological illustration of BFI by 
\citet{Jonker2025BFI}.

For the software demonstration, the data are divided into a list containing
one data frame per hospital:

\begin{CodeInput}
R> local_data <- split(Nurses, Nurses$hospital)
R> centers <- sort(unique(Nurses$hospital))
R> (L <- length(centers))
\end{CodeInput}
\vspace{-2.2em}
\begin{CodeOutput}
[1] 25
\end{CodeOutput}

The call to \code{split()} is used only to reproduce a federated setting on a
single computer. In an actual multicenter analysis, each element of
\code{local_data} would instead reside at its own center and would never be
constructed at the central server.

\subsection{Local analysis}
\label{sec:workflow_local}

The same model specification and prior structure are used consistently across
centers. Corresponding covariates have the same meaning, coding, reference
levels, and parameter names. For clarity and reproducibility, we also use the
same covariate ordering in all local analyses in this example. This is a
workflow convention rather than a strict requirement of the standard
aggregation step, because \code{bfi()} can align differently ordered
\code{theta_hat} vectors and \code{A_hat} matrices using their parameter
names when the same set of parameters is present \citep{BFIpackage}.

For this example, gender and ward type are explicitly represented as
two-level factors with the same levels in each center. A zero-mean Gaussian
prior with diagonal precision $0.01$ is used for the model parameters. The
following function contains all computations that are performed locally:

\begin{CodeInput}
R> local_analysis <- function(dat) {
+     X <- data.frame(age = dat$age,
+         gender = factor(dat$gender, levels = c(0, 1)),
+         experience = dat$experien,
+         wardtype = factor(dat$wardtype, levels = c(0, 1)))
+     Lambda <- inv.prior.cov(X = X, lambda = 0.01, family = "gaussian")
+     fit <- MAP.estimation(y = dat$stress, X = X, family = "gaussian",
+         Lambda = Lambda, control = list(maxit = 1000))
+     stopifnot(fit$convergence == 0)
+     list(theta_hat = fit$theta_hat, A_hat = fit$A_hat, Lambda = fit$Lambda)
+ }
\end{CodeInput}

The key point is that the object returned by \code{local_analysis()} contains
only the quantities required for the standard BFI aggregation. The response
vector and covariate matrix are not included.

For the reproducible example, we apply the same local analysis to all 25
hospitals:

\begin{CodeInput}
R> local_results <- lapply(local_data, local_analysis)
\end{CodeInput}

Conceptually, the preceding command represents 25 independent executions of \\
\code{MAP.estimation()}, one at each hospital. In practice, these calls may be
performed on different computers and at different times. 
For standard BFI model estimation, each center communicates its local MAP estimate
$\widehat{\boldsymbol{\theta}}_{\ell}$, curvature matrix
$\widehat{\mathbf{A}}_{\ell}$, and prior information to the central server
\citep{Jonker2024BFI,BFIpackage}.

Before turning to the central stage, it is useful to see what a single local
analysis returns. The model is therefore refitted for the first hospital and
summarized. In addition to the estimates, standard deviations and
credible intervals, the summary reports the model formula, the value of the log
posterior at the mode, and the convergence code. Of these quantities, only the
MAP estimate and the curvature matrix are communicated to the central server.

\begin{CodeInput}
R> fit_one <- MAP.estimation(y = local_data[[1]]$stress,
+     X = data.frame(
+         age = local_data[[1]]$age,
+         gender = factor(local_data[[1]]$gender, levels = c(0, 1)),
+         experience = local_data[[1]]$experien,
+         wardtype = factor(local_data[[1]]$wardtype, levels = c(0, 1))),
+     family = "gaussian", Lambda = local_results[[1]]$Lambda,
+     control = list(maxit = 1000))
R> summary(fit_one)
\end{CodeInput}
\vspace{-1em}
\begin{CodeOutput}
Summary of the local model:

   Formula: y ~ age + gender + experience + wardtype 
    Family: ‘gaussian’ 
      Link: ‘identity’

Coefficients:

            Estimate Std.Dev CI 2.5% CI 97.5%
(Intercept)   6.6647  0.6289  5.4320   7.8973
age          -0.0206  0.0209 -0.0616   0.0204
gender1      -0.2633  0.3152 -0.8811   0.3544
experience   -0.0048  0.0419 -0.0869   0.0773
wardtype1    -0.2054  0.3140 -0.8208   0.4101

Dispersion parameter (sigma2):  0.8335 
            log Lik Posterior:  -29.91 
                  Convergence:  0 
\end{CodeOutput}

\subsection{Preparing the information for the central server}
\label{sec:workflow_transfer}

At the central server, the local MAP estimates and curvature matrices are
collected in separate lists:

\begin{CodeInput}
R> theta_hats <- lapply(local_results, `[[`, "theta_hat")
R> A_hats <- lapply(local_results, `[[`, "A_hat")
\end{CodeInput}

In this example, the same prior precision matrix is used in every center and
for the fictive combined analysis. When all these precision matrices are
identical and \code{stratified = FALSE}, \code{bfi()} allows a single matrix
to be supplied rather than a list containing $L+1$ identical matrices
\citep{BFIpackage}. We therefore use

\begin{CodeInput}
R> Lambda <- local_results[[1]]$Lambda
\end{CodeInput}

The objects \code{theta_hats}, \code{A_hats}, and \code{Lambda} constitute
the input needed at the central server for the present homogeneous BFI
analysis. Individual-level observations are no longer involved in the
calculation.

\subsection{Central BFI aggregation}
\label{sec:workflow_central}

The local inference results are aggregated by

\begin{CodeInput}
R> fit_bfi <- bfi(theta_hats = theta_hats, A_hats = A_hats, Lambda = Lambda,
+     family = "gaussian")
\end{CodeInput}

The resulting object has class \code{"bfi"} and can be summarized using the
usual \code{summary()} interface:

\begin{CodeInput}
R> class(fit_bfi)
\end{CodeInput}
\vspace{-1em}
\begin{CodeOutput}
[1] "bfi"
\end{CodeOutput}

\begin{CodeInput}
R> summary(fit_bfi)
\end{CodeInput}
\vspace{-1em}
\begin{CodeOutput}
Summary of the BFI model:

    Family: ‘gaussian’ 
      Link: ‘identity’

Coefficients:

            Estimate Std.Dev CI 2.5% CI 97.5%
(Intercept)   5.6317  0.0805  5.4740   5.7894
age           0.0214  0.0027  0.0161   0.0268
gender1      -0.4919  0.0435 -0.5772  -0.4067
experience   -0.0625  0.0055 -0.0733  -0.0517
wardtype1    -0.0102  0.0384 -0.0854   0.0650

Dispersion parameter (sigma2):  0.5168
\end{CodeOutput}

The object \code{fit_bfi$theta_hat} contains
$\widehat{\boldsymbol{\theta}}_{\mathrm{BFI}}$, including the regression
coefficients and, for the Gaussian family, the estimated residual variance.
The object \code{fit_bfi$A_hat} contains
$\widehat{\mathbf{A}}_{\mathrm{BFI}}$, and \code{fit_bfi$sd} contains the
posterior standard deviations derived from its inverse \citep{BFIpackage}.

The curvature matrix itself can also be displayed:

\begin{CodeInput}
R> summary(fit_bfi, cur_mat = TRUE)
\end{CodeInput}

\subsection{Comparison with a pooled analysis}
\label{sec:workflow_pooled}

The primary target of BFI is the analysis that would have been performed had
the individual-level data been available for pooling. Because the
\code{Nurses} data are included in the package, this otherwise fictive
analysis can be carried out here and used as a software validation benchmark.
The pooled analysis is not part of the BFI procedure.

First, the design matrix for the complete data set is constructed using
exactly the same variable definitions and ordering as in the local analyses:

\begin{CodeInput}
R> X_pooled <- data.frame(age = Nurses$age,
+     gender = factor(Nurses$gender, levels = c(0, 1)),
+     experience = Nurses$experien,
+     wardtype = factor(Nurses$wardtype, levels = c(0, 1)))
\end{CodeInput}

The same prior specification is then used for the pooled MAP analysis:

\begin{CodeInput}
R> Lambda_pooled <- inv.prior.cov(X = X_pooled, lambda = 0.01,
+     family = "gaussian")
R> fit_pooled <- MAP.estimation(y = Nurses$stress, X = X_pooled,
+     family = "gaussian", Lambda = Lambda_pooled,
+     control = list(maxit = 1000))
\end{CodeInput}

The BFI and pooled estimates can be compared directly. The \code{vcov()}
method returns covariance matrices on the parameter scale returned by
\code{coef()}; for Gaussian models, the residual-variance component is
transformed from the \(\log(\sigma^2)\) scale to the original
\(\sigma^2\) scale using the delta method. Thus, no additional transformation
is required before comparing posterior standard deviations.

\begin{CodeInput}
R> bfi_est <- coef(fit_bfi)
R> pooled_est <- coef(fit_pooled)
R> sd_bfi <- sqrt(diag(vcov(fit_bfi)))
R> sd_pooled <- sqrt(diag(vcov(fit_pooled)))
R> comparison <- data.frame(parameter = names(bfi_est), BFI = bfi_est,
+     pooled = pooled_est, difference = bfi_est - pooled_est,
+     BFI_sd = sd_bfi, pooled_sd = sd_pooled)
R> print(comparison, digits = 3, row.names = FALSE)
\end{CodeInput}
\vspace{-1.1em}
\begin{CodeOutput}
  parameter     BFI  pooled difference  BFI_sd pooled_sd
(Intercept)  5.6317  5.4603    0.17134 0.08047   0.12460
        age  0.0214  0.0190    0.00245 0.00275   0.00424
    gender1 -0.4919 -0.4936    0.00170 0.04351   0.06690
 experience -0.0625 -0.0571   -0.00539 0.00549   0.00845
  wardtype1 -0.0102  0.0728   -0.08304 0.03838   0.05902
     sigma2  0.5168  0.8701   -0.35333 0.02312   0.03891
\end{CodeOutput}

Alternatively, the largest absolute discrepancy between the parameter
estimates can be obtained by

\begin{CodeInput}
R> max(abs(bfi_est - pooled_est))
\end{CodeInput}
\vspace{-1.5em}
\begin{CodeOutput}
[1] 0.3533337
\end{CodeOutput}

and the posterior standard deviations can be compared similarly:

\begin{CodeInput}
R> max(abs(sd_bfi - sd_pooled))
\end{CodeInput}
\vspace{-1.5em}
\begin{CodeOutput}
[1] 0.04413881
\end{CodeOutput}

\subsection{Interpretation of the example}
\label{sec:workflow_interpretation}

The preceding example deliberately uses the simplest BFI specification, in
which all parameters are assumed to be common across the 25 hospitals.
However, the \code{Nurses} data were generated from a multilevel structure
and exhibit several forms of between-hospital heterogeneity
\citep{Jonker2025BFI}. In particular, hospital size and other hospital-level
characteristics can affect the outcome distribution, and the residual
variation may also differ across hospitals.

Consequently, this example should primarily be interpreted as a demonstration
of the basic software workflow rather than as an assumption that the
homogeneous model is the most appropriate model for these data. Previous
analyses of the \code{Nurses} data showed that, when heterogeneity is ignored,
the BFI estimates of several regression coefficients are close to the
corresponding pooled estimates, whereas larger discrepancies can occur for
the common intercept and residual variance \citep{Jonker2025BFI}. This is
expected when a common-parameter model is used for data generated from
heterogeneous centers.

This feature is useful pedagogically: the comparison with the pooled
analysis illustrates that BFI is intended to reproduce the corresponding
pooled \emph{model}, not to remove model misspecification or unmodeled
between-center heterogeneity. The next section therefore extends the same
workflow to BFI models that explicitly accommodate heterogeneity across
centers.

\bigskip

\section{Handling heterogeneity across centers}
\label{sec:heterogeneity}

The homogeneous analysis in Section~\ref{sec:workflow} assumes that the
parameters of the Gaussian regression model are common across all hospitals.
In practice, multicenter data may exhibit several forms of heterogeneity.
For the \code{Nurses} data, for example, previous analyses identified
variation in mean stress levels across hospitals, an association between
hospital size and stress, and substantial variation in the locally estimated
residual variances \citep{Jonker2025BFI}.

The BFI methodology allows selected parameters of the aggregated model to
vary across centers or groups of centers. In \pkg{BFI}, these models are
specified through the arguments \code{stratified}, \code{strat_par}, and
\code{center_spec}. Importantly, the local models themselves are fitted in the
usual way. When \code{inv.prior.cov()} is used for a local analysis,
\code{stratified} must remain \code{FALSE}. The argument
\code{stratified = TRUE} is used only when constructing the prior precision
matrix for the fictive combined model and when performing the corresponding
central aggregation with \code{bfi()} \citep{BFIpackage}.

Consequently, the objects \code{theta_hats} and \code{A_hats} obtained in
Section~\ref{sec:workflow_local} can be reused for the heterogeneous analyses
below. The individual-level data do not need to be analyzed again for these
alternative central parameterizations \citep{Jonker2025BFI}.

\subsection{Center-specific parameters}
\label{sec:center_specific_parameters}

For Gaussian regression, the current implementation distinguishes two
parameters that can be made center-specific through \code{strat_par}.
Setting \code{strat_par = 1} gives center-specific intercepts,
\code{strat_par = 2} gives center-specific residual variances, and
\code{strat_par = c(1, 2)} allows both quantities to vary across centers
\citep{BFIpackage}.

\subsubsection{Center-specific intercepts}
\label{sec:center_specific_intercepts}

Differences in outcome means across centers that are not explained by the
included covariates can be represented by center-specific intercepts. In the
\code{Nurses} example this results in 25 hospital-specific intercepts instead
of one common intercept, while the regression coefficients and residual
variance remain common across hospitals \citep{Jonker2025BFI}.

The prior precision matrix for the fictive combined model is constructed by

\begin{CodeInput}
R> Lambda_intercept <- inv.prior.cov(X = X_pooled, lambda = 0.01,
+     family = "gaussian", stratified = TRUE, strat_par = 1, L = L)
\end{CodeInput}

The local prior matrices remain unchanged. Because the same local prior
precision matrix was used in all hospitals in Section~\ref{sec:workflow},
\code{bfi()} permits the prior specification to be supplied as a list of two
matrices: the first represents the common local prior and the second the prior
for the fictive combined model \citep{BFIpackage}. Hence,

\begin{CodeInput}
R> fit_intercept <- bfi(theta_hats = theta_hats, A_hats = A_hats,
+     Lambda = list(Lambda, Lambda_intercept), family = "gaussian",
+     stratified = TRUE, strat_par = 1)
R> summary(fit_intercept)
\end{CodeInput}

The estimated intercepts vary considerably across hospitals:

\begin{CodeInput}
R> est <- coef(fit_intercept)
R> round(range(est[grep("Intercept", names(est))]), 4)
\end{CodeInput}
\vspace{-1.1em}
\begin{CodeOutput}
[1] 3.9819 6.1451
\end{CodeOutput}

The resulting parameter vector contains one intercept for each hospital,
followed by the common regression parameters and the common residual variance.
The estimated intercepts range from 3.98 to 6.15, which indicates substantial
variation in mean stress levels between hospitals that is not explained by the
nurse-level covariates. This corresponds to the center-specific-intercept
model considered for the \code{Nurses} data by \citet{Jonker2025BFI}.

\subsubsection{Center-specific residual variances}
\label{sec:center_specific_variances}

Heterogeneity may also occur in a nuisance parameter. For Gaussian regression,
\pkg{BFI} allows the residual variance $\sigma^2$ to vary across centers by
setting \code{strat_par = 2}. This possibility is relevant for the
\code{Nurses} data because the locally estimated residual variances were found
to vary substantially across hospitals \citep{Jonker2025BFI}.

The combined prior precision matrix is now constructed as

\begin{CodeInput}
R> Lambda_sigma <- inv.prior.cov(X = X_pooled, lambda = 0.01,
+     family = "gaussian", stratified = TRUE, strat_par = 2, L = L)
\end{CodeInput}

and the central aggregation becomes

\begin{CodeInput}
R> fit_sigma <- bfi(theta_hats = theta_hats, A_hats = A_hats,
+     Lambda = list(Lambda, Lambda_sigma), family = "gaussian",
+     stratified = TRUE, strat_par = 2)
R> summary(fit_sigma)
\end{CodeInput}

The estimated residual variances vary considerably across hospitals:

\begin{CodeInput}
R> est <- coef(fit_sigma)
R> s2 <- est[grep("sigma2_loc", names(est))]
R> round(range(s2), 4)
\end{CodeInput}
\vspace{-1.1em}
\begin{CodeOutput}
[1] 0.1670 1.1629
\end{CodeOutput}

In this model, the intercept and regression coefficients are shared across
hospitals, whereas a separate residual variance is estimated for each
hospital. The estimated residual variances range from 0.17 to 1.16, a ratio of
approximately seven, which is consistent with the substantial variation in
locally estimated residual variances reported for these data by
\citet{Jonker2025BFI}.

In this example, allowing the residual variance to vary across centers
left the posterior standard deviations of the regression coefficients
unchanged to the reported precision, while the coefficient estimates
changed only marginally.

\subsubsection{Center-specific intercepts and residual variances}
\label{sec:center_specific_both}

The two forms of heterogeneity can also be modeled simultaneously. This is
specified by

\begin{CodeInput}
R> Lambda_both <- inv.prior.cov(X = X_pooled, lambda = 0.01,
+     family = "gaussian", stratified = TRUE, strat_par = c(1, 2), L = L)
R> fit_both <- bfi(theta_hats = theta_hats, A_hats = A_hats,
+     Lambda = list(Lambda, Lambda_both), family = "gaussian",
+     stratified = TRUE, strat_par = c(1, 2))
R> summary(fit_both)
\end{CodeInput}

The aggregated model then contains hospital-specific intercepts and
hospital-specific residual variances, while the regression coefficients of the
nurse-level covariates remain common across hospitals. The estimates in this
model are close to those obtained when each form of heterogeneity is modeled
separately: the largest absolute difference is below 0.001 for the intercepts
and about 0.003 for the residual variances. This parameterization corresponds to
the most flexible center-specific model considered in the analysis of the
\code{Nurses} data by \citet{Jonker2025BFI}.

\subsection{Heterogeneity due to a center-specific variable}
\label{sec:center_specific_covariate}

A different type of heterogeneity arises when a relevant covariate is
constant within each center but varies between centers. Such a variable cannot
be estimated as an ordinary covariate in the separate local regressions,
because it has no within-center variation. Nevertheless, it can be incorporated
when the local inference results are aggregated \citep{Jonker2025BFI}.

Hospital size provides an example in the \code{Nurses} data. Each hospital is
classified as small, medium, or large, and all nurses within a hospital
therefore have the same value of this variable. The published analysis of
these data found that hospital size was associated with stress and used it as
a center-specific variable in the BFI model \citep{Jonker2025BFI}.
The levels of \code{hospsize} are coded 0, 1 and 2 for small, medium and large
hospitals, and these codes appear in the parameter names.

The hospital-size value must be collected in the same center order used for
\code{theta_hats} and \code{A_hats}. Because \code{local_data},
\code{local_results}, \code{theta_hats}, and \code{A_hats} were constructed
in the same order in Section~\ref{sec:workflow}, the vector can be obtained by

\begin{CodeInput}
R> Hsize <- vapply(local_data, function(dat) dat$hospsize[1], numeric(1))
R> Hsize <- factor(Hsize)
\end{CodeInput}

The conversion to a factor makes explicit that hospital size is used as a
categorical center-specific variable. The ordering of \code{Hsize} must match
the ordering of the centers in the local BFI input objects
\citep{BFIpackage}.

The prior precision matrix for the fictive combined model is then

\begin{CodeInput}
R> Lambda_hsize <- inv.prior.cov(X = X_pooled, lambda = 0.01,
+     family = "gaussian", stratified = TRUE, center_spec = Hsize, L = L)
\end{CodeInput}

For a \code{center_spec} analysis, \code{strat_par} remains \code{NULL}.
The model is aggregated using

\begin{CodeInput}
R> fit_hsize <- bfi(theta_hats = theta_hats, A_hats = A_hats,
+     Lambda = list(Lambda, Lambda_hsize), family = "gaussian",
+     stratified = TRUE, center_spec = Hsize)
R> summary(fit_hsize)
\end{CodeInput}
\vspace{-1.5em}
\begin{CodeOutput}
Summary of the BFI model:

    Family: ‘gaussian’ 
      Link: ‘identity’

Coefficients:

              Estimate Std.Dev CI 2.5% CI 97.5%
(Intercept)_0   5.1143  0.0907  4.9365   5.2920
(Intercept)_1   5.5956  0.0820  5.4350   5.7562
(Intercept)_2   6.0644  0.0890  5.8900   6.2388
age             0.0221  0.0027  0.0167   0.0275
gender1        -0.4819  0.0435 -0.5672  -0.3966
experience     -0.0624  0.0055 -0.0731  -0.0516
wardtype1      -0.0088  0.0384 -0.0840   0.0664

Dispersion parameter (sigma2):  0.5171 
\end{CodeOutput}

Rather than estimating 25 separate hospital intercepts, this analysis groups
hospitals according to the levels of \code{Hsize}. The resulting model has
hospital-size-specific intercepts together with common regression
coefficients and a common residual variance. Thus, information from hospitals
belonging to the same hospital-size category contributes to the estimation of
the corresponding category-specific intercept
\citep{Jonker2025BFI,BFIpackage}.

\subsection{Summary of stratification options}
\label{sec:stratification_summary}

The five Gaussian models considered above differ only in the parameterization
of the fictive combined model. The homogeneous analysis uses
\code{stratified = FALSE}. Center-specific intercepts, center-specific residual
variances, or both are obtained with \code{stratified = TRUE} and
\code{strat_par} equal to \code{1}, \code{2} or \code{c(1, 2)}. A center-level
covariate is supplied instead through \code{center_spec}, in which case
\code{strat_par} remains \code{NULL}.

% Table~\ref{tab:stratification_options} summarizes the heterogeneous Gaussian
% models illustrated above. In all cases, the local MAP estimates and curvature
% matrices are unchanged; the difference lies in the parameterization of the
% fictive combined model and its prior precision matrix.

% \begin{table}[t!]
% \centering
% \begin{tabular}{@{}L{0.40\textwidth}lll@{}}
% \toprule
% Model & \code{stratified} & \code{strat_par} & \code{center_spec} \\
% \midrule
% Homogeneous &
% \code{FALSE} & \code{NULL} & \code{NULL} \\
% \addlinespace
% Center-specific intercept &
% \code{TRUE} & \code{1} & \code{NULL} \\
% \addlinespace
% Center-specific residual variance &
% \code{TRUE} & \code{2} & \code{NULL} \\
% \addlinespace
% Center-specific intercept and residual variance &
% \code{TRUE} & \code{c(1, 2)} & \code{NULL} \\
% \addlinespace
% Hospital-size-specific intercept &
% \code{TRUE} & \code{NULL} & \code{Hsize} \\
% \bottomrule
% \end{tabular}
% \caption{Selected heterogeneity specifications for Gaussian regression in
% \pkg{BFI}.}
% \label{tab:stratification_options}
% \end{table}

These examples illustrate an important distinction between local estimation
and central model specification in \pkg{BFI}. For the forms of heterogeneity
considered here, the local centers fit the same regression models and provide
the same MAP estimates and curvature matrices. Alternative assumptions about
which parameters are shared across centers are subsequently represented in
the central BFI model through its enlarged parameter vector and corresponding
prior precision matrix \citep{Jonker2025BFI,BFIpackage}.

\bigskip

\section{Survival models in practice}
\label{sec:survival}

The \pkg{BFI} package extends the same local-estimation and central-aggregation
principle to time-to-event outcomes. For \code{family = "survival"}, the
response supplied to \code{MAP.estimation()} has two columns: \code{time},
containing the observed follow-up time, and \code{status}, with 0 denoting
censoring and 1 denoting an event. In contrast to the Gaussian and binomial
models, a separate regression intercept is not fitted for survival models; its
role is incorporated in the parameterization of the baseline hazard
\citep{Pazira2026BFISurvival,BFIpackage}.

The package currently provides six choices through \code{basehaz}: exponential
(\code{"exp"}), Weibull (\code{"weibul"}), Gompertz (\code{"gomp"}), exponentiated polynomial (\code{"poly"}), piecewise exponential (\code{"pwexp"}), and an unspecified baseline hazard
(\code{"unspecified"}). These specifications follow the standard survival
literature: the Gompertz hazard \citep{Gompertz_2014}, the piecewise
exponential model \citep{Friedman_82}, and the Cox proportional hazards model
with an unspecified baseline \citep{Cox72}. The last option estimates the
regression coefficients from the partial likelihood, whereas the remaining
choices estimate parameters of the baseline hazard jointly with the regression
coefficients \citep{Pazira2026BFISurvival,BFIpackage}.

The computational workflow is standard when the dimension and interpretation
of the model parameters are fixed in advance. Two baseline-hazard choices,
however, require additional handling. For \code{"poly"}, locally selected
polynomial specifications may have different dimensions, and the central
server must reconcile them before aggregation. For \code{"pwexp"}, all centers
must use an identical set of time intervals so that corresponding baseline
hazard parameters have the same interpretation across centers
\citep{Pazira2026BFISurvival,BFIpackage}.

\subsection{A reproducible two-center survival example}
\label{sec:survival_data}

For illustration, we use simulated survival data. The setup below follows the
survival example used in the package documentation: three continuous
covariates are generated in each of two centers, and survival times are
generated from a Weibull model with right censoring \citep{BFIpackage}. The
function \code{surv.simulate()} generates the censoring times so that a
prespecified censoring proportion is attained, following the approach of
\citet{Martinez_2016}.

\begin{CodeInput}
R> set.seed(112358)
R> p <- 3; beta <- 1:3; a <- 5; b <- 6; n1 <- 30; n2 <- 30
R> X1 <- data.frame(matrix(rnorm(n1 * p), n1, p))
R> y1 <- surv.simulate(Z = list(X1), beta = beta, a = a, b = b, u1 = 0.1,
+     cen_rate = 0.3, gen_data_from = "weibul")$D[[1]][, 1:2]
R> X2 <- data.frame(matrix(rnorm(n2 * p), n2, p))
R> y2 <- surv.simulate(Z = list(X2), beta = beta, a = a, b = b, u1 = 0.1,
+     cen_rate = 0.3, gen_data_from = "weibul")$D[[1]][, 1:2]
\end{CodeInput}

The objects \code{y1} and \code{y2} contain only \code{time} and
\code{status}; the covariates are supplied separately through \code{X1} and
\code{X2}. As in the other model families, the covariate definitions and
factor reference categories must represent the same model in every center
\citep{BFIpackage}.

\subsection{Fixed-dimensional baseline hazards}
\label{sec:survival_fixed}

For exponential, Weibull, Gompertz, and unspecified baseline hazards, the
local parameter vectors have a fixed dimension once the covariate specification
and baseline-hazard family have been chosen. Consequently, the usual
\code{theta_hat}/\code{A_hat} workflow can be used. For example, a homogeneous
Weibull BFI analysis is obtained as follows
\citep{Pazira2026BFISurvival,BFIpackage}.

\begin{CodeInput}
R> Lambda_weib <- inv.prior.cov(X = X1, lambda = c(0.1, 1),
+     family = "survival", basehaz = "weibul")
R> fit_weib1 <- MAP.estimation(y = y1, X = X1, family = "survival",
+     Lambda = Lambda_weib, basehaz = "weibul")
R> fit_weib2 <- MAP.estimation(y = y2, X = X2, family = "survival",
+     Lambda = Lambda_weib, basehaz = "weibul")
R> fit_weib_bfi <- bfi(
+     theta_hats = list(fit_weib1$theta_hat, fit_weib2$theta_hat),
+     A_hats = list(fit_weib1$A_hat, fit_weib2$A_hat),
+     Lambda = Lambda_weib, family = "survival", basehaz = "weibul")
R> summary(fit_weib_bfi)
\end{CodeInput}
\vspace{-1.5em}
\begin{CodeOutput}
Summary of the BFI model:

    Family: ‘survival’ 
  Baseline: ‘weibul’

Coefficients:

        Estimate Std.Dev CI 2.5% CI 97.5%
X1        1.3653  0.2283  0.9179   1.8127
X2        2.1354  0.2992  1.5490   2.7217
X3        3.5779  0.4515  2.6931   4.4628
omega_1   1.6714  0.2822  1.1183   2.2244
omega_2   1.9004  0.1232  1.6589   2.1419
\end{CodeOutput}

With a two-element \code{lambda} vector for a survival model, the first value
specifies the prior precision for the regression coefficients and the second
specifies the prior precision for the baseline-hazard parameters. For the
Weibull and Gompertz models, two baseline-hazard parameters are estimated in
addition to the regression coefficients \citep{BFIpackage}.

\begin{figure}[t!]
\centering
\includegraphics[width=0.65\textwidth]{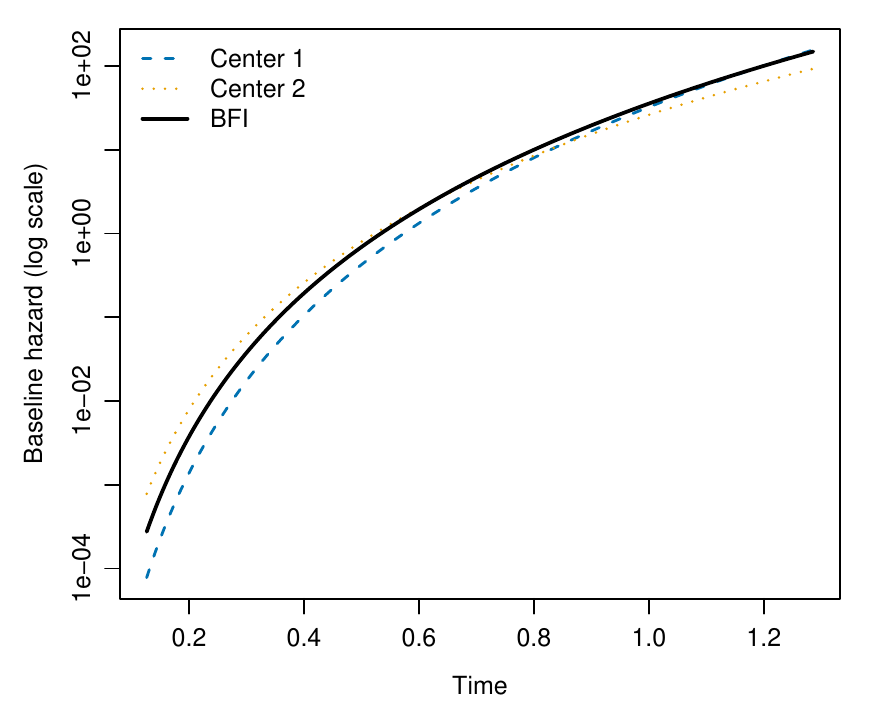}
\caption{Estimated Weibull baseline hazard functions from the two local
analyses and from the BFI aggregation, shown on a logarithmic scale over the
range of the observed follow-up times. The figure illustrates the software
workflow and is not intended as a comparison of statistical performance.}
\label{fig:weibull_basehaz}
\end{figure}

The estimated baseline hazard can be evaluated with \code{hazards.fun()}.
Figure~\ref{fig:weibull_basehaz} shows the baseline hazard implied by the two
local analyses and by the aggregated model, over the range covered by the
observed follow-up times. With 30 observations per center, the two local
estimates differ noticeably. The aggregated estimate is not a pointwise
average of the two local curves: in this example, it lies outside the
pointwise range spanned by the two local estimates at 48\% of the evaluated
time points, forming a single contiguous segment from approximately
\(t=0.63\) to \(t=1.18\). This can occur because BFI aggregates the model
parameters rather than the hazard function itself, and the mapping from these
parameters to the baseline hazard is nonlinear.

\subsection{Exponentiated polynomial baseline hazard}
\label{sec:survival_poly}

The exponentiated polynomial model represents the logarithm of the baseline
hazard by a polynomial, which can be motivated as a truncated series expansion
of the log baseline hazard \citep{Taylor2013}. In the software, a model indexed by \code{q_l}
contains the baseline-hazard parameters
$\omega_0,\ldots,\omega_{q_l}$ and therefore contains \code{q_l + 1}
baseline-hazard coefficients. The local polynomial specification can be
selected by a sequence of likelihood-ratio tests controlled by the arguments
\code{alpha} and \code{max_order} \citep{Pazira2026BFISurvival,BFIpackage}.

It is important that \code{max_order} is common across centers. The prior
precision matrix is constructed for the largest candidate model because
\code{MAP.estimation()} may need to evaluate polynomial specifications up to
that value. The defaults are \code{alpha = 0.1} and
\code{max_order = 2}; here they are specified explicitly to make the
federated protocol visible in the code \citep{BFIpackage}.

\begin{CodeInput}
R> alpha_poly <- 0.1; max_order <- 2
R> Lambda_poly <- inv.prior.cov(X = X1, lambda = c(0.1, 1),
+     family = "survival", basehaz = "poly", max_order = max_order)
R> fit_poly1 <- MAP.estimation(y = y1, X = X1, family = "survival",
+     Lambda = Lambda_poly, basehaz = "poly", alpha = alpha_poly,
+     max_order = max_order)
R> fit_poly2 <- MAP.estimation(y = y2, X = X2, family = "survival",
+     Lambda = Lambda_poly, basehaz = "poly", alpha = alpha_poly,
+     max_order = max_order)
\end{CodeInput}

Each call returns the locally selected value \code{q_l}. In addition,
\code{theta_A_poly} contains the MAP estimates and curvature matrices needed
for candidate specifications from the locally selected \code{q_l} through
\code{max_order}. Thus, the local center already computes the information
that may subsequently be required if another center selects a larger
polynomial specification \citep{BFIpackage}.

\begin{CodeInput}
R> (q_ls <- c(fit_poly1$q_l, fit_poly2$q_l))
\end{CodeInput}
\vspace{-1.5em}
\begin{CodeOutput}
[1] 2 2
\end{CodeOutput}
\vspace{-1.5em}
\begin{CodeInput}
R> (q_max <- max(q_ls))
\end{CodeInput}
\vspace{-1.5em}
\begin{CodeOutput}
[1] 2
\end{CodeOutput}

In this realization, both centers select \code{q_l = 2}, so that the
combined polynomial index is also \code{q_max = 2}.
The central aggregation is different from the standard BFI call. In
particular, the local \code{theta_hat} and \code{A_hat} objects must not be
supplied to \code{bfi()}. Instead, the complete \code{theta_A_poly} arrays and
the selected \code{q_l} values are transmitted to the central server
\citep{BFIpackage}.

\begin{CodeInput}
R> theta_A_polys <- list(fit_poly1$theta_A_poly, fit_poly2$theta_A_poly)
R> fit_poly_bfi <- bfi(Lambda = Lambda_poly, family = "survival",
+     basehaz = "poly", theta_A_polys = theta_A_polys, q_ls = q_ls)
R> summary(fit_poly_bfi)
\end{CodeInput}
\vspace{-1em}
\begin{CodeOutput}
Summary of the BFI model:

    Family: ‘survival’ 
  Baseline: ‘poly’

Coefficients:

        Estimate Std.Dev CI 2.5% CI 97.5%
X1        0.6377  0.1918  0.2618   1.0135
X2        0.7019  0.1531  0.4018   1.0019
X3        1.3883  0.2131  0.9705   1.8060
omega_0  -1.6270  0.3536 -2.3201  -0.9339
omega_1   3.6309  0.7166  2.2263   5.0354
omega_2  -0.4161  0.3617 -1.1250   0.2928
\end{CodeOutput}

Internally, \code{bfi()} sets the combined polynomial index to
\[
q_{\max} = \max_{\ell=1,\ldots,L} q_\ell.
\]
It then extracts from each local \code{theta_A_poly} object the MAP estimate
and curvature matrix corresponding to this common specification before
applying the BFI aggregation. Consequently, centers may select different
local polynomial specifications without requiring another local analysis
after the results have been communicated to the central server
\citep{Pazira2026BFISurvival,BFIpackage}.

This implementation detail is particularly important for preserving the
one-shot character of BFI. Conceptually, selection of the largest local
polynomial specification determines which common model should be aggregated.
Computationally, however, the potentially required higher-order local fits
have already been stored during the original call to
\code{MAP.estimation()}, so the central server does not need to request a
refit from the local centers \citep{Pazira2026BFISurvival,BFIpackage}.

\subsection{Piecewise exponential baseline hazard}
\label{sec:survival_pwexp}

The piecewise exponential model represents the baseline hazard by a constant
hazard within each of a prespecified number of time intervals \citep{Friedman_82}. If
\code{n_intervals = K}, the model contains $K$ baseline-hazard parameters,
one for each interval. A fundamental requirement in the federated setting is
that these parameters refer to the same intervals in every center; otherwise,
parameters with different meanings would be combined at the central server
\citep{Pazira2026BFISurvival,BFIpackage}.

The current \pkg{BFI} implementation uses a common \code{n_intervals} and a common scalar \code{min_max_times} across centers.
Each center first determines the maximum of its observed time variable, after
which the central server defines \code{min_max_times} as the minimum of these
center-specific maxima. The resulting common value is then supplied to every
local \code{MAP.estimation()} call, so that the piecewise-exponential baseline
hazard is parameterized using the same interval definition in all centers
\citep{BFIpackage}.

For the two-center example, the required preliminary summary is

\begin{CodeInput}
R> n_intervals <- 3
R> local_max_times <- c(max(y1$time), max(y2$time))
R> (min_max_times <- min(local_max_times))
\end{CodeInput}
\vspace{-1.5em}
\begin{CodeOutput}
[1] 1.286064
\end{CodeOutput}

Only these scalar maximum-time summaries are required for this setup step;
the individual survival times remain local. After the common value has been
determined, it is used by both centers together with the same number of
intervals \citep{BFIpackage}.

\begin{CodeInput}
R> Lambda_pwexp <- inv.prior.cov(X = X1, lambda = c(0.1, 1), 
+     family = "survival", basehaz = "pwexp", n_intervals = n_intervals)
R> fit_pwexp1 <- MAP.estimation(y = y1, X = X1, family = "survival",
+     Lambda = Lambda_pwexp, basehaz = "pwexp", n_intervals = n_intervals,
+     min_max_times = min_max_times)
\end{CodeInput}
\vspace{-1.7em}
\begin{CodeOutput}
No. observations in the intervals :  10 11 9 
\end{CodeOutput}
\begin{CodeInput}
R> fit_pwexp2 <- MAP.estimation(y = y2, X = X2, family = "survival",
+     Lambda = Lambda_pwexp, basehaz = "pwexp", n_intervals = n_intervals,
+     min_max_times = min_max_times)
\end{CodeInput}
\vspace{-1.7em}
\begin{CodeOutput}
No. observations in the intervals :  14 7 5 
\end{CodeOutput}

Once the common intervals have been fixed, the piecewise-exponential model
returns ordinary \code{theta_hat} and \code{A_hat} objects. Its central
aggregation therefore uses the standard BFI interface, unlike the
exponentiated-polynomial case \citep{BFIpackage}.

\begin{CodeInput}
R> fit_pwexp_bfi <- bfi(
+     theta_hats = list(fit_pwexp1$theta_hat, fit_pwexp2$theta_hat),
+     A_hats = list(fit_pwexp1$A_hat, fit_pwexp2$A_hat),
+     Lambda = Lambda_pwexp, family = "survival", basehaz = "pwexp")
R> summary(fit_pwexp_bfi)
\end{CodeInput}
\vspace{-1.4em}
\begin{CodeOutput}
Summary of the BFI model:

    Family: ‘survival’ 
  Baseline: ‘pwexp’

Coefficients:

        Estimate Std.Dev CI 2.5% CI 97.5%
X1        0.7173  0.2035  0.3184   1.1162
X2        0.7903  0.1681  0.4609   1.1196
X3        1.6146  0.2535  1.1179   2.1114
omega_1  -1.8333  0.4075 -2.6321  -1.0346
omega_2   0.7332  0.2637  0.2163   1.2500
omega_3   2.0341  0.4394  1.1729   2.8953
\end{CodeOutput}

An important practical distinction is that \code{bfi()} itself does not
receive \code{n_intervals} or \code{min_max_times}. Agreement on the
piecewise-exponential intervals is therefore part of the analysis protocol
that precedes local estimation rather than something that can be reconstructed
or verified during central aggregation. Once this common parameterization has
been established, only the local MAP estimates and curvature matrices are
needed for the final BFI aggregation \citep{BFIpackage}.

\subsection{Information flow for the survival models}
\label{sec:survival_information_flow}

Table~\ref{tab:survival_workflows} summarizes the main differences between
the survival workflows. The exponentiated-polynomial model requires richer
local output because the centrally selected polynomial specification is not
known before the local analyses. The piecewise-exponential model instead
requires harmonization of the interval definition before local estimation
\citep{Pazira2026BFISurvival,BFIpackage}.

\begin{table}[t!]
\centering
\begin{tabular}{@{}L{0.22\textwidth}L{0.23\textwidth}L{0.24\textwidth}%
L{0.21\textwidth}@{}}
\toprule
Baseline hazard &
Information produced locally &
Input used by \code{bfi()} &
Additional coordination \\
\midrule
Exponential, Weibull, Gompertz &
\code{theta_hat}, \code{A_hat} &
\code{theta_hats}, \code{A_hats} &
Common model specification \\
\addlinespace
Exponentiated polynomial &
\code{q_l}, \code{theta_A_poly} &
\code{q_ls}, \code{theta_A_polys} &
Common \code{alpha} and \code{max_order} \\
\addlinespace
Piecewise exponential &
\code{theta_hat}, \code{A_hat} &
\code{theta_hats}, \code{A_hats} &
Common \code{n_intervals} and \code{min_max_times} \\
\addlinespace
Unspecified baseline &
\code{theta_hat}, \code{A_hat} &
\code{theta_hats}, \code{A_hats} &
Common Cox model specification \\
\bottomrule
\end{tabular}
\caption{Quantities produced in each center, supplied to \code{bfi()}, and
agreed upon in advance, for the survival models implemented in \pkg{BFI}.}
\label{tab:survival_workflows}
\end{table}

Thus, the distinction between the two flexible parametric baseline hazards
occurs at different stages of the federated workflow. For
\code{"poly"}, the local centers can select different specifications and the
central server reconciles them from the richer objects already computed
locally. For \code{"pwexp"}, the interval structure must instead be
harmonized before the local MAP estimates are obtained
\citep{Pazira2026BFISurvival,BFIpackage}.

The numerical results in this section are reported to illustrate the software
outputs and the distinct computational workflows associated with the different
baseline-hazard specifications; they are not intended as a comparison of the
statistical performance of these survival models.

\bigskip

\section{Treatment-effect estimation}
\label{sec:treatment_effect}

Treatment-effect estimation in \pkg{BFI} extends the standard model-fitting
workflow described in the previous sections. The implementation supports
Gaussian, binomial, and survival outcomes. For observational data, treatment
assignment probabilities are not assumed to be known and the analysis therefore
requires two rounds of communication between the local centers and the central
server. For randomized studies, known treatment-assignment probabilities can
instead be supplied directly, so that the propensity-score estimation round is
not required \citep{BFIpackage}.

\subsection{Observational-data workflow}
\label{sec:treatment_observational}

For observational data, the first round estimates a federated propensity-score
model. Let \(Z_{\ell i}\) denote the binary treatment indicator and
\(\mathbf{x}_{\ell i}\) the covariates used to model treatment assignment.
Within each center, \code{MAP.estimation()} estimates the coefficients of a
logistic propensity-score model. Although the outcome family of the subsequent
analysis may be Gaussian, binomial, or survival, the first-round model is
handled internally as a binomial model. The resulting local coefficient
estimates and curvature matrices are combined by \code{bfi()} to obtain the
central coefficient vector \(\widehat{\boldsymbol{\gamma}}_{\mathrm{BFI}}\)
\citep{BFIpackage}.

The central estimate \(\widehat{\boldsymbol{\gamma}}_{\mathrm{BFI}}\) is then
returned to the local centers. In the second round, each center uses this
coefficient vector to calculate individual propensity scores and the associated
weights. The weighted local analysis subsequently produces a local estimate of
the treatment coefficient and its curvature information. These local
quantities are again combined with \code{bfi()} to obtain the federated
treatment-effect estimate \(\widehat{\zeta}_{\mathrm{BFI}}\)
\citep{BFIpackage}.

For Gaussian and binomial outcomes, the second-round local analysis additionally
returns the summary information stored in \code{for_ATE}. These summaries can
be aggregated centrally to obtain the treatment-effect quantities returned in
\code{Ave_Treat}, including the inverse probability of treatment weighting
(IPTW) estimator and its normalized version (\code{wIPTW}). For survival outcomes, \code{for_ATE} and \code{Ave_Treat} are not used; the central result is the federated treatment coefficient \(\widehat{\zeta}_{\mathrm{BFI}}\) \citep{BFIpackage}.

For survival outcomes, treatment-effect estimation is implemented using an
unspecified baseline hazard. Thus, the second-round calls use
\code{basehaz = "unspecified"}, corresponding to a Cox proportional hazards
model in which the regression coefficient is estimated from a weighted partial
log-likelihood \citep{BFIpackage}.

\bigskip

\subsection{A survival treatment-effect example}
\label{sec:treatment_survival_example}

We illustrate the two-round observational-data workflow using the
\code{rotterdam} breast-cancer data distributed with the \pkg{survival}
package \citep{survivalPackage}. The data contain 2,982 patients and include
information on chemotherapy, patient and tumor characteristics, recurrence,
death, and corresponding follow-up times. Chemotherapy (\code{chemo}) is used
as the binary treatment variable in this illustration.

The documentation of \code{rotterdam} notes that, for 43 patients who died
without a recorded recurrence, the recorded death time occurs after the end of
follow-up for recurrence. We therefore use the conservative recurrence-free
survival definition provided in the \pkg{survival} documentation
\citep{survivalPackage}. This example is intended to demonstrate the software
workflow and its agreement with a corresponding pooled analysis rather than to
provide a substantive causal analysis of chemotherapy.

\begin{CodeInput}
R> library("survival")
R> data("cancer", package = "survival")
R> df <- rotterdam
R> ignore <- with(df, recur == 0 & death == 1 & rtime < dtime)
R> df$status <- with(df, ifelse(recur == 1 | ignore, recur, death))
R> df$time <- with(df, ifelse(recur == 1 | ignore, rtime, dtime)) / 
+     365.25
R> df <- df[, c("time", "status", "chemo", "year", "age", "meno",
+               "size", "grade", "nodes", "pgr", "er", "hormon")]
R> df$size <- factor(df$size)
R> df$grade <- factor(df$grade)
R> df$age  <- as.numeric(scale(df$age))
R> df$pgr  <- as.numeric(scale(df$pgr))
R> df$er   <- as.numeric(scale(df$er))
R> df$year <- as.numeric(scale(df$year))
\end{CodeInput}

As in the earlier examples, the complete data set is available here only to
construct a reproducible federated-data illustration. We randomly allocate the
patients to three pseudo-centers without stratifying the allocation by
treatment. Thus, no equality of the treatment proportions across centers is
imposed.

\begin{CodeInput}
R> L <- 3
R> set.seed(11235)
R> groups <- cut(sample(nrow(df)), breaks = L, labels = FALSE)
R> datasets <- split(df, groups)
R> table(groups, df$chemo)
\end{CodeInput}
\vspace{-1.5em}
\begin{CodeOutput}
groups   0   1
     1 805 189
     2 795 199
     3 802 192
\end{CodeOutput}

With this allocation, each pseudo-center contains 994 patients and the observed
chemotherapy proportions are approximately 19.0\%, 20.0\%, and 19.3\%,
respectively. The local outcome objects contain the recurrence-free survival
time and event indicator, whereas the remaining variables form the local
design matrices.

\begin{CodeInput}
R> X_list <- lapply(datasets, function(dat)
+         dat[, !(names(dat) %in% c("time", "status")), drop = FALSE])
R> Y_list <- lapply(datasets, function(dat)
+         dat[, c("time", "status"), drop = FALSE])
\end{CodeInput}

In the first round, chemotherapy is treated as the treatment-assignment
variable and the remaining covariates form a logistic propensity-score model.
When \code{treat_round = "first"}, this binomial model is handled internally
by \code{MAP.estimation()}, also when the eventual outcome family is survival
\citep{BFIpackage}. The local coefficient estimates and curvature matrices are
then combined at the central server.

\begin{CodeInput}
R> Lambda_r1 <- lapply(X_list, function(X)
+     inv.prior.cov(X, lambda = 0.01, family = "survival",
+         treatment = "chemo", treat_round = "first"))
R> fit_r1 <- lapply(seq_len(L), function(l)
+     MAP.estimation(Y_list[[l]], X_list[[l]], family = "survival",
+         Lambda = Lambda_r1[[l]], treatment = "chemo",
+         treat_round = "first"))
R> fitbfi_r1 <- bfi(theta_hats = lapply(fit_r1, function(x) x$theta_hat),
+     A_hats = lapply(fit_r1, function(x) x$A_hat), Lambda = Lambda_r1[[1]],
+     family = "survival", treat_round = "first")
\end{CodeInput}

The resulting central coefficient vector
\(\widehat{\boldsymbol{\gamma}}_{\mathrm{BFI}}\) is returned to the centers.
In the second round, these coefficients are used locally to calculate the
propensity scores and treatment weights. For a survival outcome the package
requires \code{basehaz = "unspecified"}; the treatment coefficient is then
estimated from a weighted Cox partial log-likelihood \citep{BFIpackage}.

\begin{CodeInput}
R> Lambda_r2 <- lapply(X_list, function(X)
+     inv.prior.cov(X, lambda = 0.01, family = "survival",
+         basehaz = "unspecified", treatment = "chemo",
+         treat_round = "second"))
R> fit_r2 <- lapply(seq_len(L), function(l)
+     MAP.estimation(Y_list[[l]], X_list[[l]], family = "survival",
+         Lambda = Lambda_r2[[l]], basehaz = "unspecified",
+         treatment = "chemo", treat_round = "second",
+         gamma_bfi = fitbfi_r1$theta_hat))
R> fitbfi_r2 <- bfi(theta_hats = lapply(fit_r2, function(x) x$theta_hat),
+     A_hats = lapply(fit_r2, function(x) x$A_hat), Lambda = Lambda_r2[[1]],
+     family = "survival", basehaz = "unspecified", treat_round = "second")
R> summary(fitbfi_r2)
\end{CodeInput}
\vspace{-1em}
\begin{CodeOutput}
Summary of the BFI model:

    Family: ‘survival’ 
  Baseline: ‘unspecified’

Coefficients:

      Estimate Std.Dev CI 2.5% CI 97.5%
chemo  -0.1984  0.0365 -0.2699  -0.1268
\end{CodeOutput}

The resulting federated treatment coefficient is \(-0.1984\), with an
estimated posterior standard deviation of \(0.0365\) and an approximate
95\% credible interval from \(-0.2699\) to \(-0.1268\). This completes the
two-round federated workflow for the survival treatment-effect analysis.
Numerical comparisons with corresponding pooled weighted Cox analyses,
including a sensitivity analysis using an alternative pseudo-center
allocation, are presented in Section~\ref{sec:validation_survival}.

Because the weighted partial likelihood is a pseudo-likelihood, this
curvature-based posterior standard deviation should not be interpreted as
a robust frequentist standard error for the weighted Cox analysis.
Section~\ref{sec:validation_survival} reports the corresponding
sandwich-based robust standard error from the pooled benchmark.

\bigskip

\subsection{Average treatment-effect outputs for Gaussian and binomial models}
\label{sec:treatment_ate}

For Gaussian and binomial outcomes, the second-round local analysis returns
additional summary statistics in \code{for_ATE}. These summaries contain the
aggregated quantities required to estimate the average treatment effect (ATE) without
transmitting the individual outcomes, treatment indicators, or propensity
scores to the central server. The local \code{for_ATE} objects are supplied to
\code{bfi()} together with the second-round MAP estimates and curvature
matrices \citep{BFIpackage}.

Let \(Z_{\ell i}\) denote the binary treatment indicator, \(Y_{\ell i}\) the
outcome, and \(e_{\ell i}\) the propensity score for individual \(i\) in
center \(\ell\). Defining
\[
S_1 =
\sum_{\ell,i}\frac{Z_{\ell i}}{e_{\ell i}},
\qquad
T_1 =
\sum_{\ell,i}\frac{Z_{\ell i}Y_{\ell i}}{e_{\ell i}},
\]
and
\[
S_0 =
\sum_{\ell,i}\frac{1-Z_{\ell i}}{1-e_{\ell i}},
\qquad
T_0 =
\sum_{\ell,i}\frac{(1-Z_{\ell i})Y_{\ell i}}
                     {1-e_{\ell i}},
\]
the current implementation returns
\[
\widehat{\mathrm{ATE}}_{\mathrm{IPTW}}
=
\frac{T_1-T_0}{N},
\]
and
\[
\widehat{\mathrm{ATE}}_{\mathrm{wIPTW}}
=
\frac{T_1}{S_1}
-
\frac{T_0}{S_0},
\]
where \(N\) is the total number of observations across centers. These values
are returned in \code{Ave_Treat} as \code{IPTW} and \code{wIPTW},
respectively \citep{BFIpackage}. 
The first of these is the Horvitz--Thompson form of the IPTW estimator \citep{HorvitzThompson1952}, in which the sum of
weighted outcomes is divided by the total sample size. The second is the
normalized, or H{\'a}jek, form \citep{Hajek1971}, in which each arm is divided
by its own sum of weights. The normalized estimator is generally preferred in
practice because it is invariant to a common rescaling of the weights and is
usually more stable when some estimated propensity scores are close to zero or
one \citep{LuncefordDavidian2004}.

To illustrate this output independently of the survival example in
Section~\ref{sec:treatment_survival_example}, we use a simple two-center
binomial example. A continuous covariate and a binary treatment variable are
generated in each center, and the binary outcome is simulated from a logistic
model.

\begin{CodeInput}
R> set.seed(112358)
R> beta_bin <- 1:3; n <- c(100, 200)
R> sim_center <- function(n_l) {
+     X <- data.frame(x1 = rnorm(n_l),
+         treatment = factor(sample(1:2, n_l, replace = TRUE)))
+     eta <- beta_bin[1] + beta_bin[2] * X$x1 +
+            beta_bin[3] * as.numeric(X$treatment == "2")
+     list(X = X, y = rbinom(n_l, size = 1, prob = binomial()$linkinv(eta)))
+ }
R> centers_bin <- lapply(n, sim_center)
\end{CodeInput}

As in Section~\ref{sec:workflow}, the computations performed in each center
and at the central server are collected in two small functions. The argument
\code{round} selects the analysis round, and further arguments are passed to
\code{MAP.estimation()} or \code{bfi()}.

\begin{CodeInput}
R> local_fit <- function(d, round, ...) {
+     Lambda <- inv.prior.cov(d$X, lambda = 0.01, family = "binomial",
+         treatment = "treatment", treat_round = round)
+     MAP.estimation(d$y, d$X, family = "binomial", Lambda = Lambda,
+         treatment = "treatment", treat_round = round, ...)
+ }
R> central_fit <- function(fits, round, ...) {
+     bfi(theta_hats = lapply(fits, `[[`, "theta_hat"),
+         A_hats = lapply(fits, `[[`, "A_hat"), Lambda = fits[[1]]$Lambda,
+         family = "binomial", treat_round = round, ...)
+ }
\end{CodeInput}

For observational data, the first round estimates the federated
propensity-score model:

\begin{CodeInput}
R> fit_r1_bin <- lapply(centers_bin, local_fit, round = "first")
R> fitbfi_r1_bin <- central_fit(fit_r1_bin, round = "first")
\end{CodeInput}

The federated first-round coefficient vector is returned to the two centers
and used in their second-round weighted outcome analyses; the \code{for_ATE}
summaries created during these calls are then aggregated centrally:

\begin{CodeInput}
R> fit_r2_bin <- lapply(centers_bin, local_fit, round = "second",
+     gamma_bfi = fitbfi_r1_bin$theta_hat)
R> fitbfi_r2_bin <- central_fit(fit_r2_bin, round = "second",
+     for_ATE = lapply(fit_r2_bin, `[[`, "for_ATE"))
R> summary(fitbfi_r2_bin)
\end{CodeInput}
\vspace{-1em}
\begin{CodeOutput}
Summary of the BFI model:

    Family: ‘binomial’ 
      Link: ‘Logit’

Coefficients:

            Estimate Std.Dev CI 2.5% CI 97.5%
(Intercept)   0.4999  0.1196  0.2656   0.7343
treatment     2.6191  0.3160  1.9997   3.2384

Dispersion parameter (sigma2):  1 

Average Treatment Effect (ATE):  

         IPTW:  0.3365 
        wIPTW:  0.3364 
\end{CodeOutput}

In this reproducible example, the federated second-round treatment
coefficient is \(2.6191\), with an estimated posterior standard deviation of
\(0.3160\). On the average-treatment-effect scale, \pkg{BFI} returns an IPTW
estimate of \(0.3365\) and a normalized weighted IPTW estimate of \(0.3364\).
The regression coefficient and the two ATE summaries are distinct quantities:
the former is the treatment coefficient in the weighted binomial regression
model, whereas the latter summarize the treatment effect on the outcome scale.
The two are nevertheless consistent. On the probability scale, the estimated
intercept and treatment coefficient imply a risk difference of \(0.335\),
close to the normalized weighted IPTW estimate.

The same second-round information flow is used for Gaussian
outcomes. For survival outcomes, \code{for_ATE} is not calculated and
\code{Ave_Treat} is not used; the central result is instead the federated
treatment coefficient described in
Section~\ref{sec:treatment_survival_example} \citep{BFIpackage}.

\bigskip

\subsection{Randomized studies with known treatment probabilities}
\label{sec:treatment_randomized}

When treatment-assignment probabilities are known by design, the federated
propensity-score estimation round is unnecessary. The \pkg{BFI} interface
allows these known probabilities to be supplied directly through
\code{RCT_propens}. In this setting, \code{gamma_bfi} is not supplied and the
analysis proceeds directly to the weighted treatment analysis
\citep{BFIpackage}.

Although the software interface uses
\code{treat_round = "second"} for this analysis, there is only one actual
analysis round in the randomized-study setting. The value \code{"second"}
identifies the treatment-analysis stage in which the known propensity scores
are used; it does not imply that a preceding propensity-score estimation round
has been performed \citep{BFIpackage}.

We illustrate this using the same two-center binomial data generated in
Section~\ref{sec:treatment_ate}. Suppose that treatment was assigned
independently with probability 0.5 for every individual. 
The known assignment probabilities are supplied directly to the two local
analyses, and the central aggregation is the same as the second-round
aggregation for observational data:

\begin{CodeInput}
R> fit_rct <- lapply(centers_bin, function(d) 
+         local_fit(d, round = "second", RCT_propens = rep(0.5, nrow(d$X))))
R> fitbfi_rct <- central_fit(fit_rct, round = "second",
+     for_ATE = lapply(fit_rct, `[[`, "for_ATE"))
R> summary(fitbfi_rct)
\end{CodeInput}
\vspace{-1.1em}
\begin{CodeOutput}
Summary of the BFI model:

    Family: ‘binomial’ 
      Link: ‘Logit’

Coefficients:

            Estimate Std.Dev CI 2.5% CI 97.5%
(Intercept)   0.3769  0.1278  0.1264   0.6274
treatment     2.9070  0.3199  2.2801   3.5340

Dispersion parameter (sigma2):  1 

Average Treatment Effect (ATE):  

         IPTW:  0.6 
        wIPTW:  0.3714 
\end{CodeOutput}

Because the treatment-assignment probabilities are known in a randomized
study, the propensity-score estimation round is omitted. Although the
software interface uses \code{treat_round = "second"}, this is the only
analysis round in the randomized-study workflow: the known assignment probabilities are
supplied through \code{RCT_propens}, while \code{gamma_bfi} remains
\code{NULL} \citep{BFIpackage}.

In this reproducible example, the federated treatment coefficient is
\(2.9070\), while the IPTW and normalized weighted IPTW estimates are
\(0.6000\) and \(0.3714\), respectively. The two ATE summaries differ because
the Horvitz--Thompson estimator divides by the total sample size rather than
by the realized sums of weights. Here, 172 of the 300 individuals received
treatment, compared with 150 expected under an assignment probability of 0.5.
When the numbers of treated and control individuals differ from their expected
values, this estimator can be markedly less stable than the normalized version
\citep{LuncefordDavidian2004}. In the observational example above, where the
propensity scores were estimated from the data, the two estimates almost
coincide.
Because all weights equal 2 in this design, the weighted log-likelihood is
twice the ordinary log-likelihood, and the reported posterior standard
deviation of the treatment coefficient (\(0.3199\)) is about \(\sqrt{2}\)
times smaller than that of an unweighted analysis. As discussed in
Section~9, the standard deviations from the weighted analyses should
therefore not be used for inference.
Thus, compared with the observational-data workflow, the main
distinction is the communication pattern: no federated propensity-score
estimation and return step is required when the assignment probabilities are
known by design.

\bigskip

\section{Numerical validation against pooled analyses}
\label{sec:validation}

A central objective of BFI is to approximate the inference that would have
been obtained had the individual-level data from the participating centers
been combined before fitting the model. The approximation is constructed from
second-order expansions of the local log-posterior densities around the local
MAP estimates, and exact numerical equality with a pooled MAP analysis is
therefore not expected in finite samples \citep{Jonker2024BFI}. Previous
methodological studies evaluated this approximation extensively by simulation
and data analysis for generalized linear and survival models
\citep{Jonker2024BFI,Jonker2025BFI,Pazira2026BFISurvival}. Here we perform
smaller software-level checks using the workflows implemented in the current
version of \pkg{BFI}.

The pooled analyses in this section are used only as validation benchmarks.
They require access to the complete individual-level data and are not part of
the federated procedure.

\subsection{Gaussian regression under a homogeneous partition}
\label{sec:validation_gaussian}

The original \code{Nurses} data exhibit between-hospital heterogeneity, as
discussed in Section~\ref{sec:heterogeneity}. To obtain a cleaner validation
of the homogeneous BFI workflow, all 1,000 nurses were randomly reassigned to
the 25 centers while retaining the original center sample sizes. This follows
the homogeneous-population validation strategy used by
\citet{Jonker2025BFI}. A fixed random seed was used to make the partition
reproducible.

\begin{CodeInput}
R> set.seed(11235)
R> hospital_sizes <- as.integer(table(Nurses$hospital))
R> random_order <- sample(seq_len(nrow(Nurses)))
R> Nurses_homo <- Nurses[random_order, ]
R> Nurses_homo$hospital_random <- rep(seq_along(hospital_sizes),
+     times = hospital_sizes)
\end{CodeInput}

A Gaussian model containing age, gender, and years of experience was fitted
locally in each randomized center and aggregated with BFI. The prior precision
parameter was set to \code{lambda = 0.01}. The same model and prior were then
fitted directly to the combined data. All 25 local optimizations and the
pooled optimization converged.

\begin{table}[t!]
\centering
\begin{tabular}{lrrrrr}
\toprule
Parameter & BFI & Pooled & Difference & BFI SD & Pooled SD \\
\midrule
Intercept  &  5.5881 &  5.4970 &  0.0910 & 0.1126 & 0.1213 \\
Age        &  0.0170 &  0.0191 & -0.0021 & 0.0039 & 0.0042 \\
Gender     & -0.4745 & -0.4921 &  0.0176 & 0.0619 & 0.0670 \\
Experience & -0.0571 & -0.0574 &  0.0003 & 0.0077 & 0.0085 \\
% \midrule
\(\sigma^2\) & 0.7940 & 0.8718 & -0.0778 & 0.0355 & 0.0390 \\
\bottomrule
\end{tabular}
\caption{BFI and pooled MAP estimates and posterior standard deviations (SD)
for the randomized homogeneous partition of the \code{Nurses} data.}
\label{tab:validation_gaussian}
\end{table}

The BFI regression-coefficient estimates were close to the corresponding
MAP estimates obtained from the pooled data, as shown in
Table~\ref{tab:validation_gaussian}. The maximum absolute difference
between the regression-coefficient estimates was \(0.0910\), attained for
the intercept, and the maximum absolute difference between their posterior
standard deviations was \(0.0087\). The estimated dispersion parameters were
\(0.7940\) for BFI and \(0.8718\) for the pooled analysis. 
%These finite-sample differences are consistent with the fact that BFI reconstructs the combined-data posterior using quadratic approximations to the local log posterior densities rather than directly optimizing the pooled-data posterior \citep{Jonker2024BFI,Jonker2025BFI}.

\subsection{Logistic regression with the trauma data}

The binomial implementation is illustrated using the \code{trauma} data,
which contain mortality outcomes for 371 trauma patients grouped into
three hospital-type subsets, with sample sizes 49, 106, and 216
\citep{Jonker2024BFI,BFIpackage}. Mortality is modeled using sex, age,
Injury Severity Score (ISS), and Glasgow Coma Scale (GCS) as covariates.
Sex is coded as 0 for male and 1 for female. The continuous covariates
age, ISS, and GCS were standardized using the mean and standard deviation
of the complete data set before constructing the local data sets.

\begin{CodeInput}
R> data("trauma", package = "BFI")
R> trauma$age <- as.numeric(scale(trauma$age))
R> trauma$ISS <- as.numeric(scale(trauma$ISS))
R> trauma$GCS <- as.numeric(scale(trauma$GCS))
R> trauma$hospital <- factor(trauma$hospital)
\end{CodeInput}

The original hospital-type subsets are retained in this example. These
subsets differ in their observed mortality rates and in some patient and
trauma characteristics \citep{Jonker2024BFI}. We nevertheless fit the
same common-parameter logistic regression model in all three subsets and
use the corresponding pooled common-parameter model as the numerical
benchmark. Accordingly, this example should be interpreted as a
software-level comparison of the federated and pooled implementations of
the specified model, rather than as evidence that the underlying center
populations are homogeneous.

\begin{CodeInput}
R> centers <- levels(trauma$hospital)
R> X_list <- lapply(centers, function(l)
+   subset(trauma, hospital == l, select = c(sex, age, ISS, GCS)))
R> y_list <- lapply(centers, function(l)
+   trauma$mortality[trauma$hospital == l])
R> Lambda_list <- lapply(X_list, function(X)
+   inv.prior.cov(X, lambda = 0.01, family = "binomial"))
R> fit_local <- lapply(seq_along(centers), function(l)
+   MAP.estimation(y = y_list[[l]], X = X_list[[l]],
+     family = "binomial", Lambda = Lambda_list[[l]]))
\end{CodeInput}

A zero-mean Gaussian prior with diagonal precision
\(\lambda = 0.01\) was used in the local analyses and in the pooled
benchmark. The local MAP estimates and curvature matrices were then
aggregated using the standard binomial BFI workflow.

\begin{CodeInput}
R> X_combined <- trauma[, c("sex", "age", "ISS", "GCS")]
R> Lambda_combined <- inv.prior.cov(X_combined, lambda = 0.01,
+   family = "binomial")
R> fit_bfi_bin <- bfi(
+   theta_hats = lapply(fit_local, function(x) x$theta_hat),
+   A_hats = lapply(fit_local, function(x) x$A_hat),
+   Lambda = c(Lambda_list, list(Lambda_combined)), family = "binomial")
R> fit_pooled_bin <- MAP.estimation(y = trauma$mortality, X = X_combined,
+   family = "binomial", Lambda = Lambda_combined)
\end{CodeInput}

All three local optimizations and the pooled optimization converged.
The resulting BFI and pooled MAP estimates and posterior standard
deviations are shown in Table~\ref{tab:validation_binomial}.

\begin{table}[t]
\centering
\begin{tabular}{lrrrrr}
\hline
Parameter & BFI & Pooled & Difference & BFI SD & Pooled SD \\
\hline
Intercept & -1.4434 & -1.6255 &  0.1822 & 0.2465 & 0.2398 \\
Sex       & -0.2473 & -0.3357 &  0.0884 & 0.4187 & 0.3990 \\
Age       &  1.2189 &  1.3703 & -0.1514 & 0.2190 & 0.2056 \\
ISS       &  0.4939 &  0.5498 & -0.0560 & 0.1945 & 0.1862 \\
GCS       & -1.7375 & -1.9981 &  0.2606 & 0.2491 & 0.2379 \\
\hline
\end{tabular}
\caption{BFI and pooled MAP estimates and posterior standard deviations (SD)
for the logistic regression analysis of the \code{trauma} data.}
\label{tab:validation_binomial}
\end{table}

All corresponding regression-coefficient estimates had the same sign.
The largest absolute difference between the BFI and pooled coefficient
estimates was \(0.2606\), attained for GCS, whereas the maximum absolute
difference between their posterior standard deviations was \(0.0197\),
attained for sex. 
%These finite-sample discrepancies are consistent with the fact that BFI reconstructs the combined-data inference from local quadratic posterior approximations rather than by directly optimizing the pooled-data posterior \citep{Jonker2024BFI}. Because the original hospital-type partition is retained here, the comparison is intended as a numerical validation of the software workflow under the specified common-parameter logistic model and not as evidence of homogeneity between the three hospital populations.

\subsection{Survival treatment-effect analysis}
\label{sec:validation_survival}

The survival treatment-effect workflow in
Section~\ref{sec:treatment_survival_example} provides a third numerical
validation setting. For the main random allocation of the
\code{rotterdam} data to three pseudo-centers, the federated treatment
coefficient was \(-0.1984\). Because the complete data are available in
this reproducible example, pooled weighted Cox analyses can be used as
external numerical benchmarks. These pooled analyses are not part of
the federated workflow.

We first construct individual propensity scores from the federated
first-round coefficient vector. To retain the separate risk-set
structure of the local Cox analyses, the pooled Cox model is stratified
by pseudo-center, while a common chemotherapy coefficient is estimated.
The propensity scores are held fixed at those obtained from the
federated first-round model.

\begin{CodeInput}
R> X_ps <- model.matrix(~ year + age + meno + size + grade +
+                       nodes + pgr + er + hormon, data = df)
R> gamma_bfi <- coef(fitbfi_r1)
R> ps_bfi <- as.numeric(plogis(X_ps %*% gamma_bfi))
R> Wprop_bfi <- df$chemo / ps_bfi + (1 - df$chemo) / (1 - ps_bfi)
R> df_pooled <- df
R> df_pooled$center <- factor(groups)
R> fit_pooled_bfi_ps <- coxph(Surv(time, status) ~ chemo + strata(center),
+     data = df_pooled, weights = Wprop_bfi, robust = TRUE, ties = "breslow")
R> coef(fit_pooled_bfi_ps)["chemo"]
\end{CodeInput}
\vspace{-1.5em}
\begin{CodeOutput}
     chemo
-0.199185
\end{CodeOutput}

\begin{CodeInput}
R> summary(fit_pooled_bfi_ps)$coefficients
\end{CodeInput}
\vspace{-1.5em}
\begin{CodeOutput}
           coef exp(coef)   se(coef)  robust se         z   Pr(>|z|)
chemo -0.199185 0.8193983 0.03651121 0.08984078 -2.217089 0.02661701
\end{CodeOutput}

The pooled weighted Cox coefficient was therefore \(-0.1992\), compared
with the BFI coefficient of \(-0.1984\), an absolute difference of
approximately \(0.00081\). The point estimates agree closely. The BFI
posterior standard deviation (\(0.0365\)) also equals the model-based
standard error of the pooled fit (\(0.0365\)), whereas the robust standard
error is \(0.0898\), about 2.5 times larger. This difference reflects the
pseudo-likelihood structure of the weighted analysis rather than a
discrepancy between the federated and pooled implementations.

As a broader benchmark, a propensity-score model is fitted directly to
the complete data using logistic regression:

\begin{CodeInput}
R> fit_ps_pooled <- glm(
+     chemo ~ year + age + meno + size + grade + nodes + pgr + er + hormon,
+     data = df, family = binomial())
R> ps_pooled <- predict(fit_ps_pooled, type = "response")
R> Wprop_pooled <- df$chemo / ps_pooled + (1 - df$chemo) / (1 - ps_pooled)
R> fit_pooled <- coxph(Surv(time, status) ~ chemo + strata(center),
+     data = df_pooled, weights = Wprop_pooled, robust = TRUE,
+     ties = "breslow")
R> coef(fit_pooled)["chemo"]
\end{CodeInput}
\vspace{-1.5em}
\begin{CodeOutput}
     chemo
-0.2034198
\end{CodeOutput}

This pooled analysis gives a treatment coefficient of \(-0.2034\), with
an absolute difference of approximately \(0.00504\) from the BFI
estimate. Unlike the preceding comparison, this benchmark also includes
differences arising from estimating the propensity-score model directly
from the pooled data.

To examine sensitivity to the artificial construction of the
pseudo-centers, the analysis was repeated after randomly allocating
patients within the treated and untreated groups separately. This
produced pseudo-centers with similar chemotherapy proportions while
retaining random allocation within each treatment group. The complete
two-round BFI workflow and the corresponding pooled Cox analyses were
then repeated using this alternative center allocation.

% \begin{table}[t]
% \centering
% \begin{tabular}{lrrr}
% \hline
% Allocation & BFI & Pooled, BFI PS & Pooled, pooled PS \\
% \hline
% Random pseudo-center allocation     & -0.1984 & -0.1992 & -0.2034 \\
% Allocation within treatment groups  & -0.2004 & -0.2023 & -0.2068 \\
% \hline
% \end{tabular}
% \caption{Treatment coefficients from the federated analysis and pooled
% center-stratified Cox benchmarks under two pseudo-center allocations of the
% \code{rotterdam} data. PS, propensity score.}
% \label{tab:validation_survival}
% \end{table}

Under the alternative allocation, the BFI treatment coefficient was
\(-0.2004\), compared with \(-0.2023\) when the BFI-derived propensity scores
were used in the pooled fit and \(-0.2068\) when the propensity scores were
estimated from the pooled data. The corresponding absolute differences from the
BFI coefficient are \(0.0019\) and \(0.0065\). Together with the values
\(0.0008\) and \(0.0050\) obtained for the main allocation, all four
discrepancies are below \(0.01\) on the log hazard-ratio scale.

Taken together, the two allocations provide a sensitivity check of the
software workflow with respect to the artificial pseudo-center
construction. The benchmark using BFI-derived propensity scores is the
closer check of the second-round aggregation because the propensity
model is held fixed, whereas the pooled propensity-score analysis gives
a broader benchmark of the complete two-stage procedure. These
comparisons are intended as numerical software validation and not as a
substantive causal analysis of chemotherapy in the \code{rotterdam}
cohort.

\bigskip

\section{Discussion and limitations}
\label{sec:discussion}

% The \pkg{BFI} package provides a unified implementation of Bayesian federated
% inference for Gaussian, binomial, and survival regression models, including
% extensions for several forms of between-center heterogeneity and for
% treatment-effect estimation \citep{Jonker2024BFI,Jonker2025BFI,
% Pazira2026BFISurvival,BFIpackage}. Its main practical feature is that the
% individual-level observations remain at the centers. Standard BFI analyses
% require the local MAP estimates and corresponding curvature matrices to be
% transmitted to the central server, where they are combined without requiring
% access to the local individual-level data.

It is useful to position \pkg{BFI} relative to existing software for
decentralized analysis. \pkg{DataSHIELD} \citep{DataSHIELD} is primarily an
analysis infrastructure: it provides remote execution of analysis commands
together with disclosure control, and the statistical methods available to the
user are those exposed by the server-side packages. \pkg{distcomp}
\citep{distcomp} implements distributed model fitting through repeated
communication between a master process and worker sites, and its initial scope
is a stratified Cox model and a distributed singular value decomposition. The
\proglang{R} package \pkg{pda} \citep{pdapackage} is methodologically closest
to \pkg{BFI}: it also targets one-shot or few-shot communication and covers
linear, logistic, Poisson and Cox models, and it accommodates heterogeneity
across sites through site-specific baseline hazards \citep{Duan2020,Luo2022}.

The two approaches differ in what is communicated and in what the central
computation targets. The surrogate-likelihood algorithms implemented in
\pkg{pda} construct a local approximation to the pooled likelihood at a lead
site and are formulated in a frequentist framework. \pkg{BFI} instead combines
local MAP estimates and local curvature matrices within a Bayesian framework,
so that the prior is an explicit part of the model specification and the
aggregated curvature matrix provides the posterior standard deviations used for
credible intervals. This is an advantage when local sample sizes are small relative to the number of parameters, because the prior regularizes the local analyses. Conversely, \pkg{pda} covers Poisson regression, which \pkg{BFI} currently does not, and \pkg{DataSHIELD} provides operational disclosure control that \pkg{BFI} does not address.

An important feature of BFI is that the central estimator is based on a
second-order approximation of the local log-posterior densities around their
local MAP estimates. Consequently, BFI should not in general be interpreted
as an algebraically exact reconstruction of the pooled-data MAP analysis.
The approximation is expected to be most reliable when the local posterior
distributions are adequately represented by their local quadratic
approximations. When a local sample size is small relative to the dimension
of the parameter space, the accuracy of this approximation may deteriorate
\citep{Jonker2024BFI,Jonker2025BFI}. Regularization through the prior can help
stabilize local estimation, but the choice of prior precision also affects
the degree of shrinkage and should therefore be made as part of the model
specification \citep{Jonker2024BFI}.

Successful local optimization is also essential. The BFI construction assumes
that the required local MAP solutions and curvature matrices can be obtained.
A non-unique posterior mode prevents the standard construction from being
applied directly, and numerical optimization requires additional care in the
presence of multiple local maxima \citep{Jonker2024BFI}. The
\code{MAP.estimation()} implementation therefore reports an optimization
convergence code and provides controls for, among others, the maximum number
of iterations and the convergence tolerance \citep{BFIpackage}. As illustrated
by the \code{Nurses} analysis in this article, these diagnostics should be
checked before local summaries are aggregated.

Between-center heterogeneity is another central modeling consideration.
Differences between centers are not automatically eliminated by federated
aggregation. If parameters that are assumed to be common in the fitted model
actually differ systematically between centers, discrepancies between BFI and
a corresponding pooled-data model can result. The BFI framework and the
current package allow several structured forms of heterogeneity to be
specified explicitly, including center-specific parameters and center-level
covariates \citep{Jonker2025BFI,BFIpackage}. More general multilevel
formulations with random center effects or random slopes are not part of the
current package implementation \citep{Jonker2025BFI}.

The survival functionality involves an additional balance between model
flexibility and local sample size. Low-dimensional baseline hazards such as
the exponential, Weibull, and Gompertz forms are parsimonious but may be too
restrictive for complex hazard shapes. The exponentiated-polynomial and
piecewise-exponential alternatives provide greater flexibility, but increasing
their complexity can lead to overfitting when the local data contain limited
information \citep{Pazira2026BFISurvival}. Moreover, flexible survival
workflows require some quantities to be harmonized across centers. In the
current implementation, the polynomial workflow combines information about
the locally selected orders, whereas the piecewise-exponential workflow
requires a common number of intervals and a centrally defined
\code{min_max_times} value \citep{BFIpackage}. More general data-driven tuning of flexible baseline-hazard models, for example selection of knot locations or penalized-spline complexity
using criteria such as cross-validation or the Akaike or Bayesian information
criterion, while preserving
a one-round federated workflow, remains an area for further methodological
development \citep{Pazira2026BFISurvival}.

% The one-round character of BFI applies to the standard regression and survival
% aggregation workflows. Treatment-effect estimation for observational data has
% a different communication pattern in the current implementation. A first
% federated round estimates the propensity-score model, after which the
% federated propensity-model coefficients are returned to the centers for the
% weighted second-round outcome analysis. When treatment probabilities are
% known by design, as in a randomized study, this first round is unnecessary
% \citep{BFIpackage}.

Treatment-effect estimation for observational data requires two communication
rounds (Section~\ref{sec:treatment_effect}), and the current implementation has
two further limitations. First, the posterior standard deviations returned for the weighted second-round analysis are based on the curvature of a weighted
pseudo-likelihood and therefore do not have the usual interpretation. In the
\code{rotterdam} example of Section~\ref{sec:validation_survival}, the robust
standard error of the corresponding pooled fit is about two and a half times
the model-based one. Accordingly, intervals based on the curvature-derived
standard deviation would be substantially narrower than intervals based on
the pooled sandwich standard error. A federated counterpart of the sandwich
variance is not currently implemented. Second, the estimated
propensity scores are treated as fixed in the second round, so the uncertainty
of the first-round estimation is not propagated. The quantities returned in
\code{Ave_Treat} are point estimates and are not accompanied by standard
errors. Consistent variance estimators for inverse probability weighted treatment
effects are available in the pooled setting \citep{LuncefordDavidian2004}, and
a federated counterpart is a natural direction for future work.

A practical federated analysis also requires agreement between centers about
the statistical model and the representation of its variables. Covariates,
factor reference categories, transformations, parameter definitions, and
model specifications must have the same statistical meaning across local
analyses. The one-round strategy is particularly useful when these decisions
can be agreed upon before the analyses are run; the coordinating analyst can,
for example, distribute common analysis code to the participating centers
\citep{Jonker2025BFI}.
Participating centers do not all need to work in \proglang{R}. Because each
center runs its local analysis only once, the local stage can be executed from
another environment through the standard interfaces to \proglang{R}. The
package documentation includes vignettes showing how \code{MAP.estimation()}
can be called from \proglang{Python} \citep{Python} through \pkg{rpy2}
\citep{rpy2} and from \proglang{SAS} through \code{PROC IML}, the latter
requiring \proglang{SAS/IML} and the \code{RLANG} system option
\citep{SASIML, BFIpackage}.

Finally, the fact that individual-level observations remain local should be
distinguished from a formal privacy guarantee. The current BFI workflow
reduces the information exchanged between centers to inferential summaries,
principally parameter estimates and curvature information
\citep{Jonker2024BFI,BFIpackage}. The package does not present this mechanism
as differential privacy or as a cryptographic secure-aggregation protocol.
Consequently, requirements concerning disclosure control, secure
communication, authentication, and institutional governance remain separate
from the statistical BFI methodology and should be addressed according to the
setting in which the software is deployed.

The current package is therefore aimed primarily at federated inference for
low-dimensional statistical regression models for which uncertainty
quantification and accurate approximation of pooled-data inference are more
important than minimizing the size of the communicated summaries. In
particular, transmitting curvature matrices entails information that grows
quadratically with the number of model parameters, making the present
approach different in scope from federated-learning methods designed for
very high-dimensional models \citep{Jonker2025BFI}.

\bigskip

\section{Conclusions}
\label{sec:conclusions}

The \pkg{BFI} package provides an \proglang{R} implementation of Bayesian federated
inference for statistical regression models when individual-level data cannot
be combined across centers. The current implementation brings Gaussian,
binomial, and survival regression into a common computational framework and
extends the basic workflow to structured between-center heterogeneity and
treatment-effect estimation
\citep{Jonker2024BFI,Jonker2025BFI,Pazira2026BFISurvival,BFIpackage}.

The central design of the software separates local estimation from central
aggregation. Local centers fit the agreed statistical model and communicate
MAP estimates together with curvature information, while the central
\code{bfi()} step combines these inferential summaries. For the standard BFI
workflow, this avoids repeated cycling between the participating centers and
retains the individual-level observations at their original locations
\citep{Jonker2024BFI,Jonker2025BFI}. The resulting estimator is designed to
approximate the inference that would have been obtained from the corresponding
combined-data analysis, with its finite-sample accuracy depending on the
quality of the local posterior approximations
\citep{Jonker2024BFI}.

The purpose of this article is therefore not to introduce a new
methodological variant of BFI, but to make the methodology developed
in the accompanying statistical literature accessible through a
consistent and reproducible software interface. 
The examples in this article demonstrate the complete
information flow from local model fitting to central aggregation, including
prior specification, heterogeneous models, alternative survival baseline
hazards, treatment-effect workflows, and comparisons with pooled-data
benchmarks
\citep{Jonker2024BFI,Jonker2025BFI,Pazira2026BFISurvival,BFIpackage}.

\section*{Computational details}

The results in this paper were obtained using \proglang{R}~4.5.1 \citep{R}
on 64-bit Windows~11 with the \pkg{BFI}~3.2.0 \citep{BFIpackage} and
\pkg{survival}~3.8-3 \citep{survivalPackage} packages. \proglang{R} itself
and all packages used are available from the Comprehensive \proglang{R}
Archive Network (CRAN) at \url{https://CRAN.R-project.org/}. Because the
local analyses rely on numerical optimization, results obtained on other
platforms may differ in the last reported digit. A single commented
replication script reproducing all results reported in this article is
provided as supplementary material.

% \section*{Acknowledgments}

% We thank the editors and the anonymous reviewers for their constructive suggestions.

\bibliography{bfibib}

\end{document}